\documentclass[11pt,a4paper]{article}
\usepackage{amsmath}
\usepackage{amsfonts}
\usepackage{amssymb}
\usepackage{amsthm}
\usepackage{mathrsfs}
\usepackage{dsfont}
\usepackage{paralist}
\usepackage{natbib}
\usepackage{bm}
\usepackage{color}
\usepackage[colorlinks=true,linkcolor=blue,citecolor=blue,pdfborder={0 0 0}]{hyperref}
\usepackage{graphicx}
\usepackage{tabularx}
\usepackage[left=3.5cm,right=3.5cm,top=2cm,bottom=2cm,includeheadfoot]{geometry}
\usepackage{comment}

\newcount\Comments  
\newcommand{\kibitz}[2]{\ifnum\Comments=1\textcolor{#1}{#2}\fi}

\newtheoremstyle{normal}
{2ex}               
{3ex}               
{}                  
{}                  
{\bfseries} 
{}                  
{2pt}   
{\thmname{#1}\thmnumber{ #2.} \thmnote{(#3)}}

\newtheoremstyle{italic}
{2ex}
{3ex}
{\itshape}
{}
{\bfseries} 
{}
{2pt}
{\thmname{#1}\thmnumber{ #2.} \thmnote{(#3)}}

\theoremstyle{normal}
\newtheorem{definition}{Definition}[section]
\newtheorem{remark}[definition]{Remark}

\newtheorem{condition}[definition]{Condition}

\theoremstyle{italic}
\newtheorem{theorem}[definition]{Theorem}
\newtheorem{lemma}[definition]{Lemma}
\newtheorem{proposition}[definition]{Proposition}
\newtheorem{corollary}[definition]{Corollary}

\newcommand\N{\mathbb{N}}

\newcommand\R{\mathbb{R}}

\newcommand\eps{\varepsilon}
\newcommand\Prob{\mathbb{P}}    
\newcommand\Exp{\mathbb{E}}     

\newcommand\Bb{\mathbb{B}}

\newcommand\Eb{\mathbb{E}}

\newcommand\Gb{\mathbb{G}}

\newcommand\Tb{\mathbb{T}}

\newcommand\Vb{\mathbb{V}}

\DeclareMathOperator{\Cov}{Cov}

\newcommand\weak{\rightsquigarrow}
\newcommand\weakP{{\, {\weak_\xi}\ }}
\newcommand\pto{\stackrel{\scriptscriptstyle \mathbb P^*}{\rightarrow}}
\newcommand\defeq{:=}
\newcommand{\ip}[1]{\lfloor #1 \rfloor}

\allowdisplaybreaks[1]

\begin{document}

\title{Nonparametric tests for detecting breaks in the jump behaviour of a time-continuous process
}


\author{Axel B\"ucher\footnotemark[1]\ ,  Michael Hoffmann\footnotemark[1]\  ,  Mathias Vetter\footnotemark[2] ~and Holger Dette\footnotemark[1]\ , \bigskip \\
{ Ruhr-Universit\"at Bochum \ \& \ Philipps-Universit\"at Marburg}
}

\footnotetext[1]{Ruhr-Universit\"at Bochum,
Fakult\"at f\"ur Mathematik, 44780 Bochum, Germany.
{E-mail:} axel.buecher@rub.de, holger.dette@rub.de, michael.hoffmann@rub.de}

\footnotetext[2]{Philipps-Universit\"at Marburg, Fachbereich Mathematik und Informatik, 35032 Marburg, Germany.
{E-mail:} vetterm@mathematik.uni-marburg.de}


\maketitle

\begin{abstract}
This paper is concerned with tests for changes in the jump behaviour of a time-continuous process. Based on results on weak convergence of a sequential empirical tail integral process, asymptotics of certain tests statistics for breaks in the jump measure of an It\=o semimartingale are constructed. Whenever limiting distributions depend in a complicated way on the unknown jump measure, empirical quantiles are obtained using a multiplier bootstrap scheme. An extensive simulation study shows a good performance of our tests in finite samples.
\end{abstract}


\noindent \textit{Keywords and Phrases:}  Change points; L\'evy measure; multiplier bootstrap; sequential empirical processes; weak convergence.

\smallskip

\noindent \textit{AMS Subject Classification:} 60F17, 60G51, 62G10.


\section{Introduction}
\def\theequation{1.\arabic{equation}}
\setcounter{equation}{0}

Recent years have witnessed a growing interest in statistical tools for high-frequency observations of time-continuous processes. With a view on finance, the seminal paper by \cite{DelSch94} suggests to model such a process using It\=o semimartingales, say $X$, which is why most research has focused on the estimation of (or on tests concerned with) its characteristics. Particular interest has been paid to integrated volatility or the entire quadratic variation, mostly adapting parametric procedures based on normal distributions, as the continuous martingale part of an It\^o semimartingale is nothing but a time-changed Brownian motion. For an overview on methods in this field see the recent monographs by \cite{JacPro12} and \cite{AitJac14}.

Still less popular is inference on the jump behaviour only, even though empirical research shows a strong evidence supporting the presence of a jump component within $X$; see e.g.\ \cite{AitJac09a} or \cite{AitJac09b}. 
In this work, we will address the question whether the jump behaviour of $X$ is time-invariant. Corresponding tests, commonly referred to as change point tests, are well known in the framework of discrete time series, but have recently also been extended to time-continuous processes; see e.g.\ \cite{LeeNisYos06} on changes in the drift or \cite{IacYos12} on changes in the volatility function of $X$. However, to the best of our knowledge, no procedures are available for detecting breaks in the jump component.

Suppose that we observe an It\=o semimartingale $X$ which admits a decomposition of the form
\begin{multline} \label{ItoSem}
X_t = X_0 +  \int_0^t b_s \, ds + \int_0^t \sigma_s\, dW_s + \int_0^t \int_{\R} u 1_{\{|u| \leq 1\}} (\mu-\bar \mu)(ds,du) \\
+ \int_0^t \int_{\R} u 1_{\{|u|>1\}} \mu(du,dz),
\end{multline}
where $W$ is a standard Brownian motion, $\mu$ is a Poisson random measure on $\R^+ \times \R$, and the predictable compensator $\bar \mu$ satisfies $\bar \mu(ds,du) = ds \: \nu_s(du)$. 
As a fairly general structural assumption, we allow the characteristics of $X$, i.e.\ $b_t,\sigma_t$ and $\nu_t,$ to depend deterministically on time.  Recall that $\nu_t$ can be interpreted as a local L\'evy measure, such that
\[
\int_{\R} (1 \wedge |u|^2) \nu_t(du) < \infty
\]
for each $t$ and $\nu_t(A)$ denotes the average number of jumps that fall into the set $A$ over a unit time interval.


Now, we assume that we have data from the process in a high-frequency setup. Precisely, at stage $n \in \mathbb N$, we are able to
observe realizations of the process $X$ at the equidistant times $i \Delta_n$ for $i = 1, \ldots, n$, where the mesh $\Delta_n \rightarrow 0$, while $n \Delta_n \rightarrow \infty$. In this situation we want to test the null hypothesis that the jump behaviour of the process is the same for all $n$ observations, i.e.\ there exists some measure $\nu$ such that $\nu_t(dz) = \nu(dz)$ for all $t$, against alternatives involving the non-constancy of $\nu_t$. For instance, one might consider an alternative consisting of one break point, i.e.\ there exists some $\theta_0 \in (0,1)$ and two L\'evy measures $\nu_1$, $\nu_2$  such that the process behind the first $\lfloor n\theta_0 \rfloor$ observations has L\'evy measure $\nu_1$ and the remaining $n - \lfloor n\theta_0 \rfloor$ observations are taken from a process with L\'evy measure $\nu_2$.
The restriction to a deterministic drift and volatility in \eqref{ItoSem} is merely technical here, as it allows to use empirical process theory for independent observations later. An argument similar to that in Section 5.3 in \cite{BueVet13} proves that one might as well work with random coefficients $b$ and $\sigma$.

Throughout the work, we will restrict ourselves to positive jumps only. Thus, for $z>0$, let $U(z) \defeq \nu([z, \infty))$ denote the tail integral (or spectral measure; see \citealp{RueWoe02}) associated with $\nu$, which determines the jump measure uniquely. For $\ell_1,\ell_2 \in \{1, \dots, n\}$ such that $\ell_1 < \ell_2$, define
\[
U_{\ell_1:\ell_2} (z)  \defeq \frac{1}{(\ell_2 - \ell_1 +1) \Delta_n} \sum \limits_{j=\ell_1}^{\ell_2} \mathtt 1_{\lbrace \Delta_j^n X \geq z \rbrace} \quad (z >0),
\]
with $\Delta_j^n X \defeq X_{j \Delta_n} - X_{(j-1) \Delta_n}$, which serves as an empirical tail integral based on the increments $\Delta_{\ell_1}^nX,  \dots, \Delta_{\ell_2}^n X$.
If $X$ is a L\'evy process with a L\'{e}vy measure $\nu$ not changing in time, \cite{FigLop08} illustrated that  $U_{1:n}(z)$ is a suitable estimator for the tail integral $U(z)$ in the sense that, under regularity conditions, $U_n(z)$ is $L^2$-consistent for $U(z)$. Following the approach in \cite{Ino01}, it is therefore likely that we can base tests for $H_0$ on suitable functionals of the process
\begin{align*}
D_n(\theta,z)
 \defeq
U_{1:\lfloor n\theta \rfloor}(z) - U_{(\lfloor n\theta \rfloor+1):n}(z),
\end{align*}
where $\theta \in [0,1]$ and $z>0$. Under the null hypothesis, this expression can be expected to converge to $0$ for all $\theta \in [0,1]$ and $z>0$, whereas under alternatives, for instance those involving a change at $\theta_0$ as described before, $D_n(\theta_0,z)$ should converge to an expression which is non-zero.

More precisely, we will consider the following standardized version of $D_n$, namely
\begin{align}
\label{Teststatistic}
\Tb_n(\theta,z) \defeq \sqrt {n\Delta_n} \lambda_n(\theta)
\left\{ U_{1:\lfloor n \theta \rfloor}(z) - U_{(\lfloor n \theta \rfloor+1):n}(z)\right \}
\end{align}
for $\theta \in [0,1]$ and $z > 0$, where $\lambda_n(\theta) = \frac{\lfloor n \theta \rfloor}{n} \frac{n - \lfloor n \theta \rfloor}{n}$. An appropriate functional
allowing to test the hypothesis of a constant L\'{e}vy measure is for instance given by a Kolmogorov-Smirnov statistic of the form
\begin{align} \label{eq:Tneps}
T_n^{(\eps)} \defeq \sup \limits_{\theta \in [0,1]} \sup \limits_{z \geq \eps} |\Tb_n( \theta,z)|, \quad (\eps > 0).
\end{align}
The null hypothesis of no change in the   L\'evy measure   is rejected for large values of $T_n^{\scriptscriptstyle (\eps)}$.
The restriction to jumps larger than $\eps$ is important, since there might be infinitely many of arbitrary small size.

The limiting distribution of the previously mentioned test statistic will turn out to depend in a complicated way on the unknown L\'evy measure~$\nu$. Therefore, corresponding quantiles are not easily accessible and must be obtained by suitable bootstrap approximations. Following related ideas for detecting breaks within multivariate empirical distribution functions \citep{Ino01}, we opt for using empirical counterparts based on a multiplier bootstrap scheme, frequently also referred to as \textit{wild} or \textit{weighted} bootstrap. The approach essentially consists of multiplying each indicator within the respective empirical tail integrals with an additional, independent and standardized multiplier. The underlying empirical process theory is for instance summarized in the monograph \cite{Kos08}.

The remaining part of this paper is organized as follows: the derivation of a functional weak convergence result for the process  $\Tb_n$  under the null hypothesis is the content of Section~\ref{sec:weak}. The  asymptotic properties  of $T_n^{\scriptscriptstyle (\eps)} $ can then  easily be derived from the continuous mapping theorem. Section~\ref{sec:boot} is concerned with the approximation of the limiting distribution using the previously described multiplier bootstrap scheme. In Section~\ref{sec:test}, we discuss the formal derivation of several tests for a time-homogeneous jump behaviour, whereas an extensive simulation study is presented in Section~\ref{sec:sim}. All proofs are deferred to the Appendix, which is Section~\ref{sec:proof}.

\section{Functional weak convergence of the sequential empirical tail integral} \label{sec:weak}
\def\theequation{2.\arabic{equation}}
\setcounter{equation}{0}

In this section, we   derive a functional weak convergence result for the process $\Tb_n$ defined in \eqref{Teststatistic}. For that purpose, we have to introduce an appropriate function space. We set $\mathbb A \defeq [0,1] \times (0, \infty)$ and let $\mathcal B_{\infty} (\mathbb A)$ denote the space of
all functions $f \colon \mathbb A \rightarrow \mathbb R$ which are bounded on every set $A \subset \mathbb A$ for which the projection onto the second coordinate, $p_2(A) = \lbrace z \in (0, \infty) \mid  \exists \theta \in [0,1] \text{ such that } (\theta,z) \in A \rbrace$, is bounded away from $0$.
Moreover, for $k \in \mathbb N$, we define $A_k \defeq [0,1] \times [k^{-1}, \infty)\subset \mathbb A$, and, for $ f,g \in \mathcal B_{\infty} (\mathbb A)$, we set
\[
d(f,g) \defeq \sum \limits_{k=1}^{\infty} 2^{-k} (\| f-g \|_{A_k} \wedge 1),
\]
where $\|f-g\|_{A_k} = \sup\{ |f(x) - g(x)| \, : \, x \in A_k\}.$
Note that $d$ defines a metric on $\mathcal B_{\infty}(\mathbb A)$ which induces the topology of uniform convergence on all sets $A$ such that its projection $p_2(A)$ is bounded away from $0$, i.e.\ a sequence of functions converges with respect to $d$ if and only if it converges uniformly on each $A_k$ \citep[][Chapter 1.6]{VanWel96}.

Furthermore to establish our results on weak convergence under the null hypothesis, we impose the following conditions.

\begin{condition}
\label{Assump1} $X$ is an It\=o semimartingale with the representation in \eqref{ItoSem} such that
\begin{enumerate}
\item[(a)] The drift $b_t$ and the volatility $\sigma_t$ are c\`agl\`ad, bounded and deterministic.
\item[(b)] There exists some L\'evy measure $\nu$ such that $\nu_t \equiv \nu$ for all $t \in \mathbb R_+$.
\item[(c)] $X$ has only positive jumps, that is, the jump measure $\nu$ is supported on $(0, \infty)$.
\item[(d)] $\nu$ is absolutely continuous with respect to the Lebesgue measure on $(0,\infty)$. Its density $h=d\nu/d\lambda$, called L\'evy density, is differentiable with derivative $h^{\prime}$ and satisfies
\[
\|h\|_{M_k} + \|h'\|_{M_k} < \infty
\]
for all $k \in \mathbb N$ with $M_k \defeq [k^{-1}, \infty)$. \qed
\end{enumerate}
\end{condition}

The next lemma is essential for the weak convergence results. Similar statements can be found in \cite{FigLopHou09}, with slightly stronger assumptions on $h$, and in \cite{BueVet13} in the bivariate case.

\begin{lemma}
\label{WahrschAbschLem}
Let $X$ be an It\=o semimartingale that satisfies Condition~\ref{Assump1}. Let further be $\delta > 0$ fixed. If $X_0 = 0$, then there exist constants $K=K(\delta) >0$ and $t_0 = t_0(\delta)>0$, depending on the bounds on the characteristics in Condition~\ref{Assump1}(a), such that the inequality
$$\left| \mathbb P (X_t \in [z, \infty)) - t \nu([z, \infty)) \right| < K t^2$$
holds for all $z \ge \delta$ and all $0<t<t_0$.
\end{lemma}

\begin{remark}
\label{WahrschAbschRem}
If $X$ is an It\=o semimartingale satisfying Condition~\ref{Assump1}, then Lem\-ma~\ref{WahrschAbschLem} implies immediately that we have
$$
\sup \limits_{s \in \mathbb R_+}
\left| \mathbb P(X_{s+t} - X_s\in [z, \infty)) - t \nu([z, \infty)) \right| \leq Kt^2
$$
as well. 
To see this note that, for each fixed $s \in \mathbb R_+$, the It\=o semimartingale $(Y^{\scriptscriptstyle (s)}_t)_{t \in \mathbb R_+}$ with $Y^{\scriptscriptstyle (s)}_t \defeq X_{s+t} - X_s$ satisfies Condition~\ref{Assump1}, and its characteristics have the same bounds as the characteristics of $X$. \qed
\end{remark}



The limiting behaviour of the process $\Tb_n$ can be deduced from the next theorem, which is a result for weak convergence of a sequential empirical tail integral process. For $\theta\in [0,1]$ and $z >0$ set
\begin{eqnarray}
\label{UnVerallgSchaetz}
U_n(\theta,z) \defeq  \frac{\ip{n\theta}}{n} U_{1:\ip{n\theta}}(z) = \frac{1}{k_n} \sum \limits_{j=1}^{\lfloor n\theta \rfloor} \mathtt 1_{\lbrace \Delta_j^n X \geq z \rbrace},
\end{eqnarray}
where $k_n \defeq n \Delta_n$ and denote  its standardized version by
\begin{equation} \label{center}
\Gb_n(\theta,z) \defeq \sqrt{k_n} \lbrace U_n(\theta,z) - \mathbb E U_n(\theta,z) \rbrace.
\end{equation}
Obviously, the sample paths of $U_n(\theta,z)$ are elements of $\mathcal B_{\infty}(\mathbb A)$.

\begin{theorem}
\label{WeakConvThm}
Let $X$ be an It\=o semimartingale that satisfies Condition \ref{Assump1}. Furthermore, assume that the observation scheme has the properties:
\[
\Delta_n \rightarrow 0, \quad \text{ } \quad n \Delta_n = k_n \rightarrow \infty.
\]
Then, 
$\Gb_n \weak \Gb$
 in $(\mathcal B_{\infty}(\mathbb A) ,d)$, where
 $\mathbb G$ is a tight mean zero Gaussian process with covariance
\[
H(\theta_1,z_1;\theta_2,z_2) \defeq \mathbb E[ \mathbb G(\theta_1, z_1) \mathbb G(\theta_2, z_2) ] = (\theta_1 \wedge \theta_2) \times \nu([z_1 \vee z_2 , \infty ))
\]
for $(\theta_1,z_1) , (\theta_2, z_2) \in \mathbb A$.
The sample paths of $\mathbb G$ are almost surely uniformly continuous on each $A_k$ $(k \in \mathbb N$) with respect to the semimetric
$$\rho(\theta_1, z_1 ; \theta_2 , z_2)
\defeq
\big\{ (\theta_1 \wedge \theta_2) \nu ( [z_1 \wedge z_2 , z_1 \vee z_2 ))  + \left| \theta_1 - \theta_2 \right|  \nu ([z_{ I (\theta_1 , \theta_2) } , \infty )) \big\}^{\frac{1}{2}}$$
with
$I(\theta_1 ,\theta_2) \defeq 1 + \mathtt 1_{\lbrace \theta_1 \leq \theta_2 \rbrace}.$
\end{theorem}

Note that we have centered $U_n(\theta,z)$ around its expectation in \eqref{center}. In most applications, however, we are interested in estimating functionals of the jump measure, and according to Lemma \ref{WahrschAbschLem} we need stronger conditions then. Precisely, we consider the process 
$$\tilde \Gb_n(\theta,z) \defeq \sqrt{k_n} \lbrace U_n(\theta,z) - \theta \nu([z, \infty)) \rbrace$$
and get, as an immediate consequence of the previous two results, the following sequential generalization of Theorem 4.2 of \cite{BueVet13}.

\begin{corollary}
\label{BueVetResVerallg}
Let $X$ be an It\=o semimartingale that satisfies Condition~\ref{Assump1}. If the observation scheme meets the conditions
\begin{align*}
\Delta_n \rightarrow 0, \quad \text{ } \quad n \Delta_n = k_n \rightarrow \infty, \quad \text{ } \quad \sqrt{k_n} \Delta_n \rightarrow 0,
\end{align*}
then $\tilde \Gb_n \weak \Gb$  in $(\mathcal B_{\infty}(\mathbb A) ,d)$, where
$\Gb$ denotes the Gaussian process from Theorem~\ref{WeakConvThm}.
\end{corollary}

A further consequence of Theorem \ref{WeakConvThm} is the desired weak convergence of the process $\Tb_n$, which was defined in \eqref{Teststatistic}, under the null hypothesis.

\begin{theorem}
\label{TnSchwKonv}
Suppose the assumptions of Corollary \ref{BueVetResVerallg} are satisfied.
Then, the process $\Tb_n$ defined in \eqref{Teststatistic} converges weakly to $\Tb$  in $(\mathcal B_{\infty}(\mathbb A) ,d)$, where
$$\mathbb T(\theta,z) = \mathbb G(\theta,z) - \theta \mathbb G(1,z)$$
for $(\theta,z) \in \mathbb A$, and where $\mathbb G$ denotes the limit process in Theorem \ref{WeakConvThm}.
$\Tb$ is a tight mean zero Gaussian process  with covariance function
$$\hat H(\theta_1,z_1;\theta_2,z_2) \defeq \mathbb E  \{ \mathbb T(\theta_1,z_1) \mathbb T(\theta_2,z_2) \} = \{ (\theta_1 \wedge \theta_2) - \theta_1 \theta_2 \}
 \nu([z_1 \vee z_2, \infty)).$$
\end{theorem}

Using the continuous mapping theorem, we are now able to derive the weak convergence of various statistics allowing for the detection of breaks in the jump behaviour. The following corollary treats the statistic $T_n^{(\eps)}$ defined in \eqref{eq:Tneps}.

\begin{corollary} \label{TnSchwKonvCor}
Under the assumptions of Corollary \ref{BueVetResVerallg} we have, for each $\eps >0$,
$$T_n^{(\eps)} \weak T^{(\eps)} := \sup \limits_{0 \leq \theta \leq 1} \sup \limits_{z \geq \eps} | \mathbb T(\theta,z)|,$$
where $\mathbb T$ is the limit process defined in Theorem \ref{TnSchwKonv}.
\end{corollary}

The covariance function of the limit process in Theorem \ref{TnSchwKonv} depends on the L\'evy measure of the underlying process, which is usually unknown in applications. If one only wants to detect changes in the tail integral of the L\'evy measure at a fixed point
$z_0$, the following proposition deals with the simple transformation
  $$\Vb_n^{(z_0)}(\theta) \defeq  \frac{\Tb_n(\theta ,z_0) }{\sqrt{U_{1:n}(z_0)}} \mathtt 1_{\lbrace U_{1:n}(z_0) >0 \rbrace}
 $$
 of $\Tb_n$ which yields a pivotal limiting distribution.

\begin{proposition}
\label{SchwKonvforz0}
Let $X$ be an It\=o semimartingale that satisfies Condition~\ref{Assump1}. Moreover, let $z_0 >0$ be a real number with $\nu([z_0, \infty)) >0$ and suppose that the underlying observation scheme meets the assumptions from Corollary~\ref{BueVetResVerallg}.
Then, $\Vb_n^{(z_0)} \weak \Bb$ in $\ell^{\infty}([0,1])$, where
 $\Bb$ denotes a standard Brownian bridge. As a consequence,
$$V_n^{(z_0)} \defeq \sup \limits_{\theta \in [0,1]} |\Vb_n^{(z_0)}(\theta)| \weak  \sup \limits_{\theta \in [0,1]} \mathbb |B(\theta)|,$$
the limiting distribution being also known as the Kolmogorov-Smirnov distribution.
\end{proposition}

\begin{remark}
We have derived the previous results under somewhat simplified assumptions on the observation scheme in order to keep the presentation rather simple. A more realistic setting could involve additional microstructure noise effects or might rely on non-equidistant data. In both cases, standard techniques still yield similar results. 

For example, in case of noisy observations, \cite{Vet14} has shown that a particular de-noising technique allows for virtually the same results on weak convergence as for the plain $U_n(\theta,z)$ in the case without noise. For non-equidistant data, the limiting covariance functions $H$ and $\hat H$ in general depend on the sampling scheme. The latter effect is well-known from high-frequency statistics in the case of volatility estimation; see e.g.\ \cite{MykZha12}. \qed
\end{remark}

%
%

\section{Bootstrap approximations for the sequential empirical tail integral} \label{sec:boot}
\def\theequation{3.\arabic{equation}}
\setcounter{equation}{0}

We have seen in Corollary \ref{TnSchwKonvCor} that the distribution of the limit $\mathbb T$ of the process $\Tb_n$ depends in a complicated way on the unknown L\'evy measure of the underlying process. However, we need the quantiles of $\mathbb T$ or at least good approximations for them to obtain a feasible test procedure. Typically, one uses resampling methods to solve this problem.

Probably the most natural way to do so is to use $U_{1:n}(z)$ in order to obtain an estimator $\hat \nu_n$ for the L\'evy measure first, and to draw a large number of independent samples of an It\=o semimartingale with L\'evy measure $\hat \nu_n$ then, possibly with estimates for drift and volatility as well. Based on each sample, one might then compute the test statistic $\Tb_n$, and by doing so one obtains empirical quantiles
 for $\mathbb T$. 
 
 However, from a computational side, such a method is computationally expensive  since one has to  generate independent It\=o
 semimartingales for each stage within the bootstrap algorithm. Therefore
 we have decided to work with an alternative bootstrap method based on multipliers, where one only needs to  generate $n$ i.i.d.\ random variables with mean zero and variance one
(see also  \citealp{Ino01}, who used a similar approach in the context of empirical processes).

Precisely, the situation now is as follows: The bootstrapped processes, say $\hat Y_n = \hat Y_n(X_1, \ldots, X_n, \xi_1, \ldots, \xi_n)$, will depend on some random variables $X_1, \ldots, X_n$ and on some random weights $\xi_1, \ldots, \xi_n$. The $X_1, \ldots, X_n$, that we consider as collected data, are defined on a probability space $(\Omega_X, \mathcal A_X, \mathbb P_X)$. The random weights $\xi_1, \ldots, \xi_n$ are defined on a distinct probability space $(\Omega_{\xi}, \mathcal A_{\xi}, \mathbb P_{\xi})$. Thus, the bootstrapped processes live on the product space
$(\Omega, \mathcal A,\mathbb P) \defeq (\Omega_X, \mathcal A_X, \mathbb P_X) \otimes (\Omega_{\xi}, \mathcal A_{\xi}, \mathbb P_{\xi})$.
The following notion of conditional weak convergence will be essential. It can be found in \cite{Kos08} on pp.\ 19--20.
\begin{definition}
\label{ConvcondDataDef}
Let $\hat Y_n = \hat Y_n (X_1, \ldots ,X_n; \xi_1, \ldots , \xi_n) \colon (\Omega, \mathcal A,\mathbb P) \rightarrow \mathbb D$ be a (bootstrapped) element in some metric space $\mathbb D$ depending on some random variables $X_1, \ldots, X_n$ and some random weights $\xi_1, \ldots, \xi_n$. Moreover, let $Y$ be a tight, Borel measurable map into $\mathbb D$. Then $\hat Y_n$ converges weakly to $Y$ conditional on the data $X_1, X_2, \ldots$ in probability, notationally $\hat Y_n \weakP Y$, if and only if
\begin{enumerate}
\item[(a)] $\sup \limits_{f \in \text{BL}_1(\mathbb D)} |\mathbb E_{\xi} f(\hat Y_n) - \mathbb E f(Y) | \pto 0,$
\item[(b)] $\mathbb E_{\xi} f( \hat Y_n)^{\ast} - \mathbb E_{\xi} f( \hat Y_n)_{\ast} \pto 0$ for all
      $f \in \text{BL}_1(\mathbb D).$
\end{enumerate}
Here, $\mathbb E_{\xi}$ denotes the conditional expectation over the weights $\xi$ given the data $X_1, \ldots, X_n$, whereas $\text{BL}_1(\mathbb D)$ is the space of all real-valued Lipschitz continuous functions $f$ on $\mathbb D$ with sup-norm $\| f \|_{\infty} \leq 1$ and Lipschitz constant $1$. Moreover, $f( \hat Y_n)^{\ast}$ and $f( \hat Y_n)_{\ast}$ denote a minimal measurable majorant and a maximal measurable minorant with respect to the joint data (including the weights $\xi$), respectively. \qed
\end{definition}

\begin{remark}
\label{rem:condweak}
$~~$
\begin{compactenum}[(i)]
\item Note that we do not use a measurable majorant or minorant in item (a) of the definition. This is justified through the fact that, in this work, all
      expressions $f(\hat Y_n)$, with a bootstrapped statistic $\hat Y_n$ and a Lipschitz continuous function $f$, are measurable functions
			of the random weights.
\item Note that the implication ``(ii) $\Rightarrow$ (i)'' in the proof of Theorem~2.9.6 in \cite{VanWel96} shows that, in general,
      conditional weak convergence $\weakP$ implies unconditional weak convergence $\weak $ with respect to the
			product measure $\mathbb P$. \qed
\end{compactenum}
\end{remark}

Throughout this paper we denote by
$$\hat \Gb_n= \hat \Gb_n(\theta,z)
=
\hat \Gb_n (X_{\Delta_n}, \ldots, X_{n \Delta_n}, \xi_1, \ldots, \xi_n;\theta,z)$$
the bootstrap approximation which is defined by
\begin{align*}
\hat \Gb_n(\theta,z)
& \defeq
\frac{1}{n \sqrt{k_n}} \sum \limits_{j=1}^{\lfloor n\theta \rfloor} \sum \limits_{i=1}^n \xi_j \lbrace \mathtt 1_{\lbrace \Delta_j^n X \geq z \rbrace} - \mathtt 1_{\lbrace \Delta_i^n X \geq z \rbrace} \rbrace \\
&=
\frac{1}{\sqrt{k_n}} \sum \limits_{j=1}^{\lfloor n\theta \rfloor} \xi_j \lbrace \mathtt 1_{\lbrace \Delta_j^n X \geq z \rbrace} -\eta_n(z) \rbrace,
\end{align*}
where $
\eta_n(z)= n^{-1} \sum_{i=1}^n \mathtt 1_{\lbrace \Delta_i^n X \geq z \rbrace}
$.
The following  theorem establishes conditional weak convergence of this bootstrap approximation for the sequential empirical tail integral process $\Gb_n$.

\begin{theorem}
\label{GnVorberKonvThm}
Let $X$ be an It\=o semimartingale that satisfies Condition~\ref{Assump1} and assume that the observation scheme meets the conditions from Theorem~\ref{WeakConvThm}.
Furthermore, let $(\xi_j)_{j \in \mathbb N}$ be independent and identically distributed random variables with mean $0$ and variance $1$, defined on a distinct probability space as described above.
Then,
$$\hat \Gb_n \weakP \mathbb \Gb$$
in $(\mathcal B_{\infty}(\mathbb A),d)$, where $\mathbb G$ denotes the limiting process of Theorem \ref{WeakConvThm}.

\end{theorem}

Theorem~\ref{GnVorberKonvThm} suggests to define the following bootstrapped counterparts of the process $\Tb_n$ defined in equation \eqref{Teststatistic}:
\begin{align*}
\hat \Tb_n (\theta,z)
&\defeq
\hat \Tb_n (X_{\Delta_n}, \ldots, X_{n \Delta_n}; \xi_1, \ldots, \xi_n; \theta,z)
\defeq
\hat \Gb_n(\theta,z) - \frac{\ip{n\theta}}{n} \hat\Gb_n(1,z) \\
&=
\sqrt {n \Delta_n} \frac{\lfloor n\theta \rfloor}{n} \frac{n - \lfloor n\theta \rfloor}{n} \bigg[ \frac{1}{\lfloor n\theta \rfloor \Delta_n} \sum
\limits_{j=1}^{\lfloor n\theta \rfloor} \xi_j \{ \mathtt 1_{\lbrace \Delta_j^n X \geq z \rbrace} - \eta_n(z) \} \\
& \hspace{3.9cm}
- \frac{1}{(n- \lfloor n\theta \rfloor) \Delta_n}
\sum \limits_{j = \lfloor n\theta \rfloor +1}^n \xi_j \{ \mathtt 1_{\lbrace \Delta_j^n X \geq z \rbrace} - \eta_n(z) \} \bigg ],
\end{align*}
The following result establishes consistency of $\Tb_n$ in the sense of Definition~\ref{ConvcondDataDef}.

\begin{theorem}
\label{BootstrTeststThm}
Under the conditions and notations of Theorem~\ref{GnVorberKonvThm}, we have
$$\hat \Tb_n \weakP \mathbb T$$
 in $(\mathcal B_{\infty}(\mathbb A),d)$, with $\mathbb T$ defined in Theorem \ref{TnSchwKonv}.
\end{theorem}

The distribution of the Kolmogorov-Smirnov-type test statistic $T_n^{(\eps)}$ defined in \eqref{eq:Tneps} can be approximated with the bootstrap statistics investigated in the following corollary. It can be proved by a simple application of Proposition 10.7 in \cite{Kos08} on an appropriate $\ell^{\infty}(A_k)$.

\begin{corollary}
Under the assumptions of Theorem \ref{GnVorberKonvThm} we have, for each $\eps > 0$,
$$
\hat T_n^{(\eps)} \defeq \sup \limits_{0 \leq \theta \leq 1} \sup \limits_{z \geq \eps} | \hat \Tb_n(\theta,z)|
\weakP
\sup \limits_{0 \leq \theta \leq 1} \sup \limits_{z \geq \eps} | \mathbb T(\theta,z)| =: T^{(\eps)}.
$$
\end{corollary}

\section{The testing procedures} \label{sec:test}
\def\theequation{4.\arabic{equation}}
\setcounter{equation}{0}

\subsection{Hypotheses}\label{subsec:hypo}

In order to derive a test procedure which utilizes the results on weak convergence from the previous two sections, we have to formulate our hypotheses first. Under the null hypothesis the jump behaviour of the process is constant. More precisely, this means the following:
\begin{enumerate}
\item[$\bf H_0$:] We observe an It\=o semimartingale as in equation \eqref{ItoSem}  with characteristic triplet $(b_t, \sigma_t, \nu)$ that satisfies Condition~\ref{Assump1}.
\end{enumerate}
We want to test this hypothesis versus the alternative that there is exactly one change in the jump behaviour. This means in detail:
%
\begin{enumerate}
\item[$\bf H_1$:] There exists some $\theta_0 \in (0,1)$ and two L\'evy measures $\nu_1 \neq \nu_2$ satisfying Con\-dition~\ref{Assump1}(c) and (d) such that, at stage $n$, we observe an It\=o semimartingale $X=X(n)$ with characteristic triplet $(b_t^{\scriptscriptstyle (n)}, \sigma^{\scriptscriptstyle (n)}_t, \nu_t^{\scriptscriptstyle (n)})$
such that
\[
\nu_t^{(n)} = \mathtt 1_{ \{ t < \ip{n\theta_0} \Delta_n \} } \nu_1 +  \mathtt 1_{\{t \ge \ip{n\theta_0} \Delta_n \} } \nu_2
\]
Furthermore, $b_t^{\scriptscriptstyle  (n)}$ and $\sigma_t^{\scriptscriptstyle  (n)}$ satisfy Condition~\ref{Assump1}(a) and are uniformly bounded in $n\in\N$ and $t>0$.
\end{enumerate}
The corresponding alternative for a fixed $z_0 >0$ is then given through:
\begin{enumerate}
\item[$\bf H_1^{(z_0)}$:] We have the situation from $\textbf H_1$, but with $\nu_1([z_0,\infty)) \neq \nu_2([z_0,\infty))$ and
$\nu_1([z_0,\infty)) \vee \nu_2([z_0,\infty)) > 0$. 
\end{enumerate}

\subsection{The tests and their asymptotic properties}

In the sequel, let $B\in\N$ be some large number and let $(\xi^{\scriptscriptstyle  (b)})_{b=1, \dots ,B}$ denote independent vectors of i.i.d.\ random variables, $\xi^{\scriptscriptstyle (b)} \defeq (\xi_j^{\scriptscriptstyle (b)})_{j=1, \dots, n}$,  with mean zero and variance one. As before, we assume that these random variables are generated independently from the original data. We denote by $\hat \Tb_{\scriptscriptstyle  n, \xi^{(b)}}$ or $\hat T^{\scriptscriptstyle  (\eps)}_{\scriptscriptstyle  n, \xi^{(b)}}$ the particular statistics calculated with respect to the data  and the $b$-th bootstrap multipliers $\xi^{\scriptscriptstyle (b)}_1, \ldots, \xi^{\scriptscriptstyle  (b)}_n$. For a given level $\alpha \in (0,1)$, we consider the following test procedures:

\begin{enumerate}
\item[\textbf{KSCP-Test1.}] Reject $\textbf H_0$ in favor of $\textbf H_1^{(z_0)}$, if $V_n^{(z_0)} \geq q_{1- \alpha}^K$, where $V_n^{(z_0)}$ is defined in Proposition~\ref{SchwKonvforz0} and where $q_{1- \alpha}^{\scriptscriptstyle K}$ denotes the $1 - \alpha$ quantile of the Kolmogorov-Smirnov-(KS-)distribution, that is the distribution of $K = \sup_{s \in [0,1]} | \mathbb B(s) |$ with a standard Brownian bridge $\mathbb B$.

\item [\textbf{KSCP-Test2.}] Reject $\textbf H_0$ in favor of $\textbf H_1^{(z_0)}$, if
\[
W_n^{(z_0)} \defeq \sup \limits_{\theta \in [0,1]} |\Tb_n(\theta,z_0)| \geq \hat q^{(B)}_{1 - \alpha}(W^{(z_0)}_n),
\]
where $\hat q^{(B)}_{1 - \alpha}(W^{(z_0)}_n)$ denotes the $(1- \alpha)$-sample quantile of $\hat W_{\scriptscriptstyle n, \xi^{(1)}}^{\scriptscriptstyle (z_0)}, \ldots, \hat W_{\scriptscriptstyle n, \xi^{(B)}}^{\scriptscriptstyle (z_0)}$, and where
$
\hat W_{\scriptscriptstyle n, \xi^{(b)}}^{\scriptscriptstyle (z_0)} \defeq \sup_{\theta \in [0,1]} | \hat \Tb_{n, \xi^{(b)}}(\theta,z_0) |.
$

\item[\textbf{CP-Test.}] Choose an appropriate small $\eps > 0$ and reject $\textbf H_0$ in favor of $\textbf H_1$, if
\[
T_n^{(\eps)} \geq \hat q^{(B)}_{1 - \alpha}(T^{(\eps)}_n),
\]
where $\hat q^{(B)}_{1 - \alpha}(T^{(\eps)}_n)$ denotes the $(1- \alpha)$-sample quantile of $\hat T_{\scriptscriptstyle n,\xi^{(1)}}^{(\eps)}, \ldots, \hat T_{\scriptscriptstyle n, \xi^{(B)}}^{(\eps)}$.
\end{enumerate}

Since $\eps > 0$ has to be chosen prior to an application of the CP-Test, we can only detect changes in the jumps larger than $\eps$. From a theoretical point of view this is not entirely satisfactory, since one is interested in distinguishing arbitrary changes in the jump behaviour. On the other hand, in most applications only the larger jumps are of particular interest, and at least the size of $\Delta_n$ provides a natural bound to disentangle jumps from volatility. Thus, a practitioner can choose a minimum jump size $\eps$ first, and use the CP-Test to decide whether there is a change in the jumps larger than $\eps$.

The following proposition shows that three aforementioned tests keep the asymptotic level $\alpha$ under the null hypothesis.

\begin{proposition} \label{CorrKSCP}
Suppose the sampling scheme meets the conditions of Corollary~\ref{BueVetResVerallg}. Then, KSCP-Test1, KSCP-Test2 and CP-Test are asymptotic level $\alpha$ tests for $\mathbf H_0$ in the sense that, under $\mathbf H_0$, for all $\alpha \in (0,1)$,
$$\lim \limits_{n \rightarrow \infty} \mathbb P(V_n^{(z_0)} \geq q^K_{1- \alpha}) = \alpha, \quad \lim \limits_{B \rightarrow \infty} \lim \limits_{n \rightarrow \infty} \mathbb P \{ W_n^{(z_0)} \geq \hat q_{1 - \alpha}^{(B)}(W_n^{(z_0)}) \}  = \alpha,$$
and
$$
\lim_{B \rightarrow \infty} \lim \limits_{n \rightarrow \infty} \mathbb P \{ T_n^{(\eps)} \geq \hat q_{1- \alpha}^{(B)}(T_n^{(\eps)}) \} = \alpha,
$$
for all $\eps >0$ such that $\nu([\eps,\infty))>0$.
\end{proposition}



The next proposition shows that the preceding tests are consistent under the fixed alternatives defined in Section~\ref{subsec:hypo}. For simplicity, we only consider alternatives involving one change point, even though the results may be extended to alternatives involving multiple breaks or even continuous changes.

\begin{proposition} \label{CorConsi}
Suppose the sampling scheme meets the conditions of Corollary~\ref{BueVetResVerallg}. Then, KSCP-Test1, KSCP-Test2 and CP-Test are consistent in the following sense: under $\mathbf H_1^{\scriptscriptstyle (z_0)}$, for all $\alpha \in (0,1)$ and all $B \in \mathbb N$, we have
$$
\lim \limits_{n \rightarrow \infty} \mathbb P(V_n^{(z_0)} \geq q_{1 - \alpha}^K)=1
\quad \text{ and } \quad
\lim \limits_{n \rightarrow \infty} \mathbb P(W_n^{(z_0)} \geq \hat q_{1- \alpha}^{(B)}(W_n^{(z_0)})) =1.
$$
Under $\mathbf H_1$, there exists an $\eps > 0$ such that, for all $\alpha \in (0,1)$ and all $B \in \mathbb N$,
$$
\lim \limits_{n \rightarrow \infty} \mathbb P \{ T_n^{(\eps)} \geq \hat q_{1-\alpha}^{(B)}(T_n^{(\eps)}) \}  = 1.$$
\end{proposition}


\subsection{Locating the change point} \label{subsec:argmax}

Let us finally discuss how to construct suitable estimators for the location of the change point. We begin with a useful proposition.

\begin{proposition}
\label{prop:tnh1}
Suppose the sampling scheme meets the conditions of Corollary~\ref{BueVetResVerallg}.  Then, under $\mathbf H_1$, $(\theta,z) \mapsto k_n^{\scriptscriptstyle -1/2} \Tb_n(\theta,z)$ converges in $\mathcal B_{\infty}(\mathbb A)$ to the function
$$
T(\theta,z) \defeq \begin{cases}
                         \theta (1-\theta_0) \{ \nu_1(z) - \nu_2(z) \} \quad \text{ if } \theta \leq \theta_0 \\
																	\theta_0 (1- \theta) \{ \nu_1(z) - \nu_2(z) \} \quad \text{ if } \theta \geq \theta_0,
																	\end{cases}$$
in outer probability,
with $\nu_1(z) \defeq \nu_1([z, \infty))$ and $\nu_2(z) \defeq \nu_2([z, \infty))$.
\end{proposition}

Since $\theta \mapsto T(\theta,z)$ attains its maximum in $\theta_0$,
natural estimators for the position of the change point are therefore given by
$$
\hat \theta_n^{(\eps)} \defeq \operatorname{arg\,max}_{\theta \in [0,1]} \sup_{z \geq \eps} |\Tb_n(\theta,z)|
$$
for the test problem $\textbf H_0$ versus $\textbf H_1$ and
$$
\tilde \theta^{(z_0)}_n \defeq \operatorname{arg\,max}_{\theta \in [0,1]} |\Tb_n(\theta,z_0)|
$$
in the setup $\textbf H_0^{(z_0)}$ versus $\textbf H_1^{(z_0)}$. The next proposition states that these estimators are consistent.

\begin{proposition} \label{CorConsiLoc}
Suppose the sampling scheme meets the conditions of Corollary~\ref{BueVetResVerallg}.
If $\mathbf H_1$ is true, there exists an $\eps > 0$ such that
$\hat \theta_n^{(\eps)} = \theta_0+ o_\Prob(1)$ as $n\to \infty$.
In the special case of $\mathbf H_1^{(z_0)}$, we have
$\tilde \theta_n^{(z_0)} = \theta_0 + o_\Prob(1).$
\end{proposition}

\section{Finite-sample performance} \label{sec:sim}
\def\theequation{5.\arabic{equation}}
\setcounter{equation}{0}

In this section, we present results of a large scale Monte Carlo simulation study, assessing the finite-sample performance of the proposed test statistics for detecting breaks in the L\'evy measure.  Moreover, under the alternative of one single break, we show results on the performance of the estimator for the break point from Section~\ref{subsec:argmax}.

The experimental design of the study is as follows. 
\begin{itemize}
\item We consider five different choices for the \textit{number of trading days}, namely $k_n=50,75,100,150,250$, and corresponding \textit{frequencies} $\Delta_n^{-1} = 450, 300, 225, 150, 90$. Note that $n=k_n \Delta_n =22,500$ for any of these choices.
\item We consider two different models for the \textit{drift} and the \textit{volatility}: either, we set $b_t =\sigma_t \equiv 0$ or $b_t =\sigma_t \equiv 1$, resulting in a pure jump process and a process including a continuous component, respectively. 
\item We consider one parametric model for the \textit{tail integral}, namely
\begin{align}
\label{eq:ubeta}
U_{\beta}(z) = \nu_{\beta}([z, \infty)) = \left(\frac{\beta}{\pi z}\right)^{1/2}, \qquad \beta >0
\end{align}
(which yields a $1/2$-stable subordinator in the case of $b_t =\sigma_t \equiv 0$).
For the parameter~$\beta$, we consider $51$ different choices, that is
$\beta = 1 + 2j/25$, with $j \in 0,\dots,50$, ranging from $\beta=1$ to $\beta=5$.
\item We consider models with one single break in the tail integral at
$50$ different \textit{break points}, ranging form $\theta_0=0$ to $\theta_0=0.98$ (note that $\theta_0=0$ corresponds to the null hypothesis). The tail integrals before and after the break point are chosen from the previous parametric model.
 \end{itemize}
 
The target values of our study are, on the one hand, the empirical rejection level of the tests and, on the other hand, the empirical distribution of the estimators for the change point $\theta_0$.
To assess these target values, any combination of the previously described settings was run $1,000$ times, with the bootstrap tests being based on $B=250$ bootstrap replications. The It\=o semimartingales were simulated by a straight-forward modification of Algorithm 6.13 in \cite{ConTan04}, where, under alternatives involving one break point, we simply merged two paths of independent semimartingales together. 

\begin{figure}[t!]
\vspace{-.2cm}
\centering
\includegraphics[width=69mm]{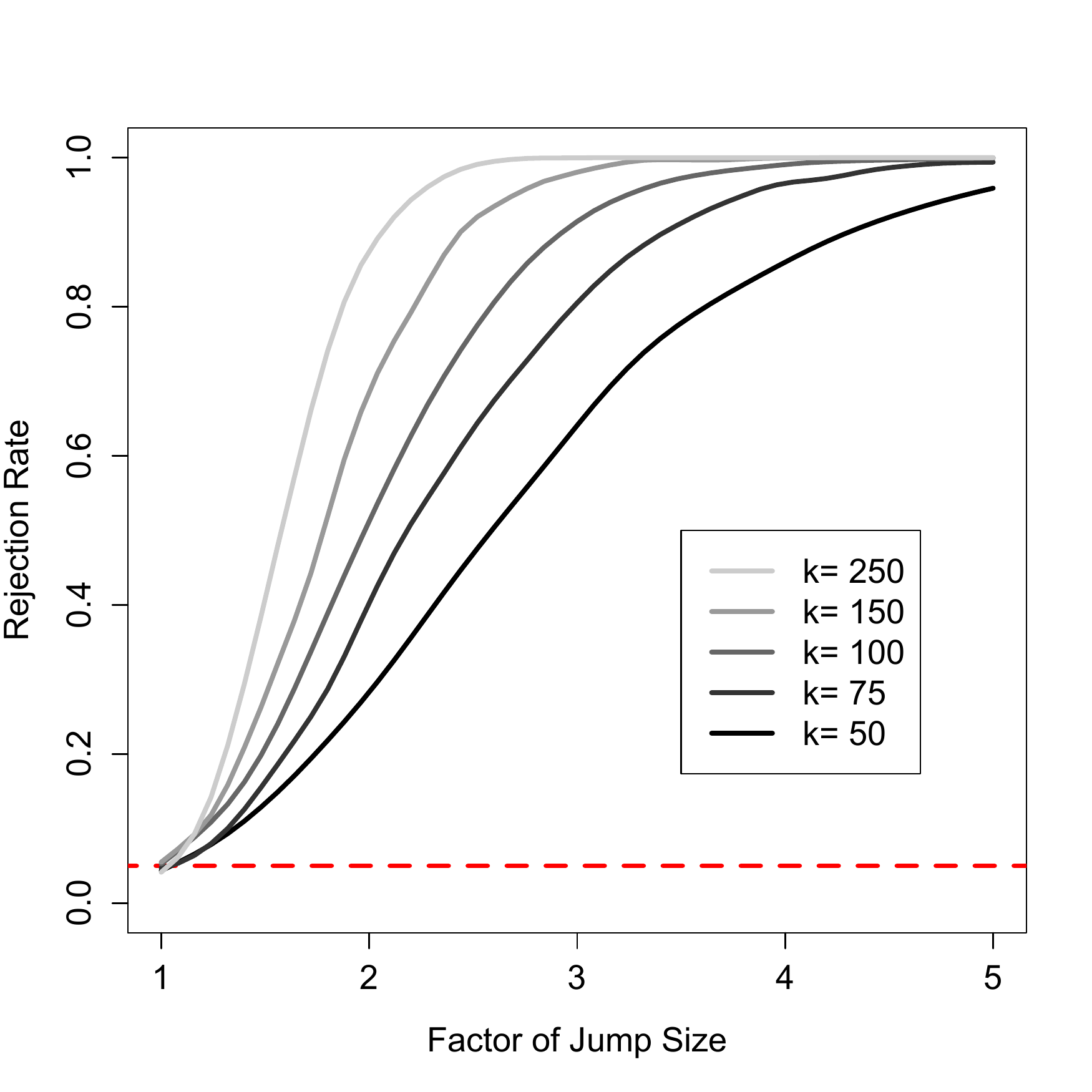}
\includegraphics[width=69mm]{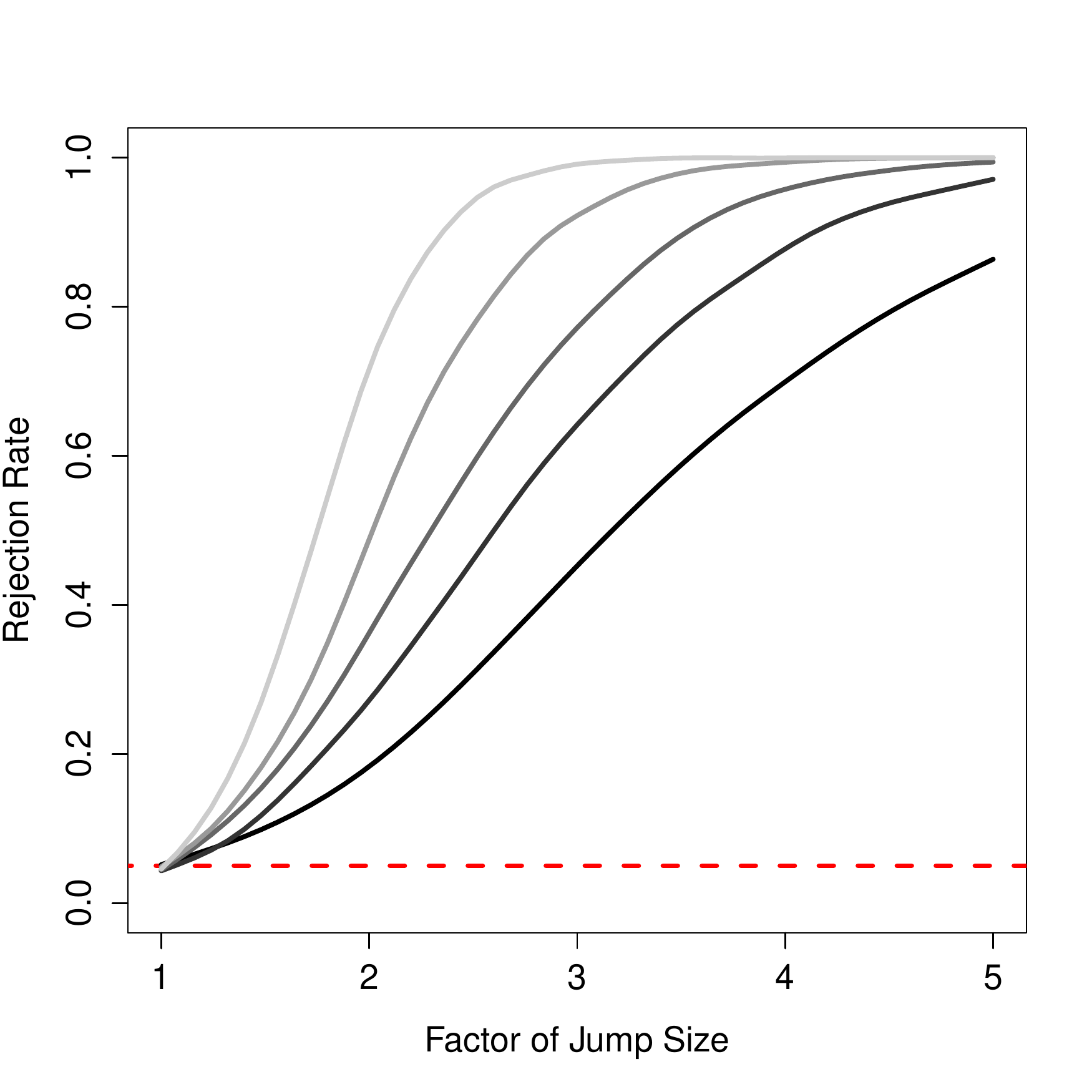}
\vspace{-.9cm}
\caption{Rejection rate of the CP-Test for pure jump subordinator data (on the left-hand side) and a subordinator plus a drift and a Brownian motion (on the right-hand side). $\beta$ changes from $1$ to the factor of jump size.}
\label{fig:h1beta}
\end{figure}

\begin{figure}[t!]
\vspace{-.2cm}
\centering
\includegraphics[width=69mm]{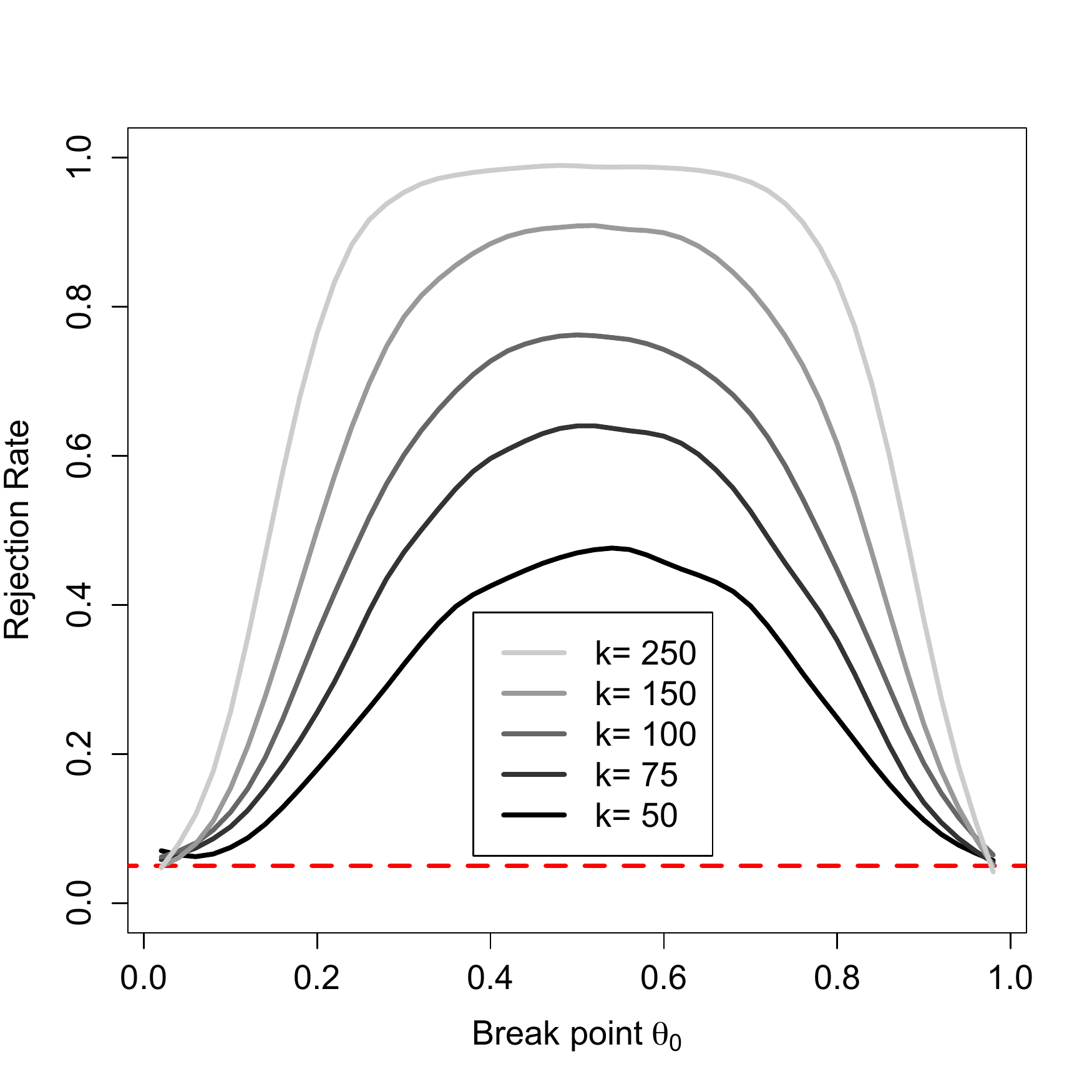}
\includegraphics[width=69mm]{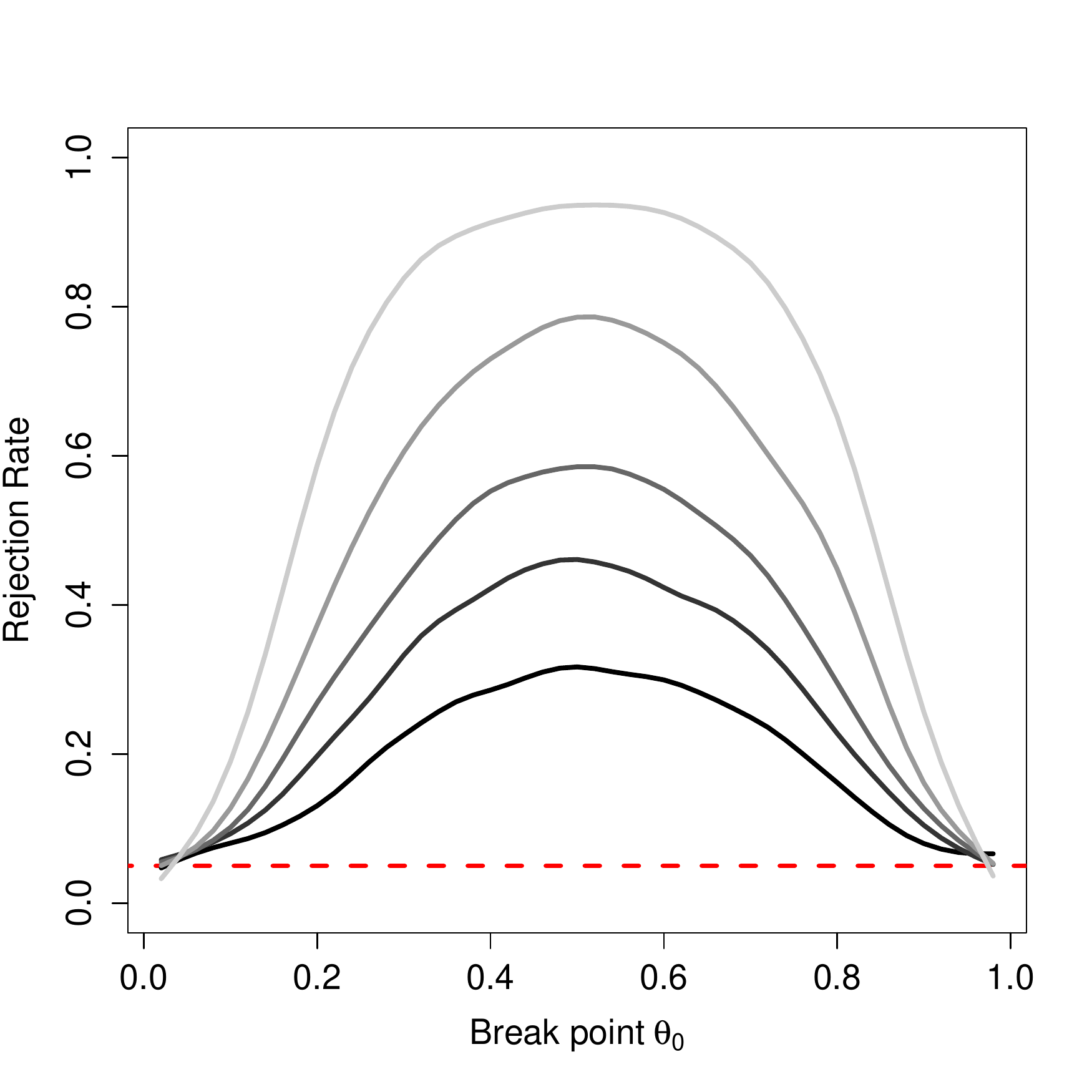}
\vspace{-.9cm}
\caption{Rejection rate of the CP-Test for pure jump subordinator data (left panel) and a subordinator with a drift plus a Brownian motion (right panel) for different change point locations.}
\label{fig:h1theta}
\end{figure}

\begin{figure}[t!]
\vspace{-.3cm}
\centering
\includegraphics[width=69mm]{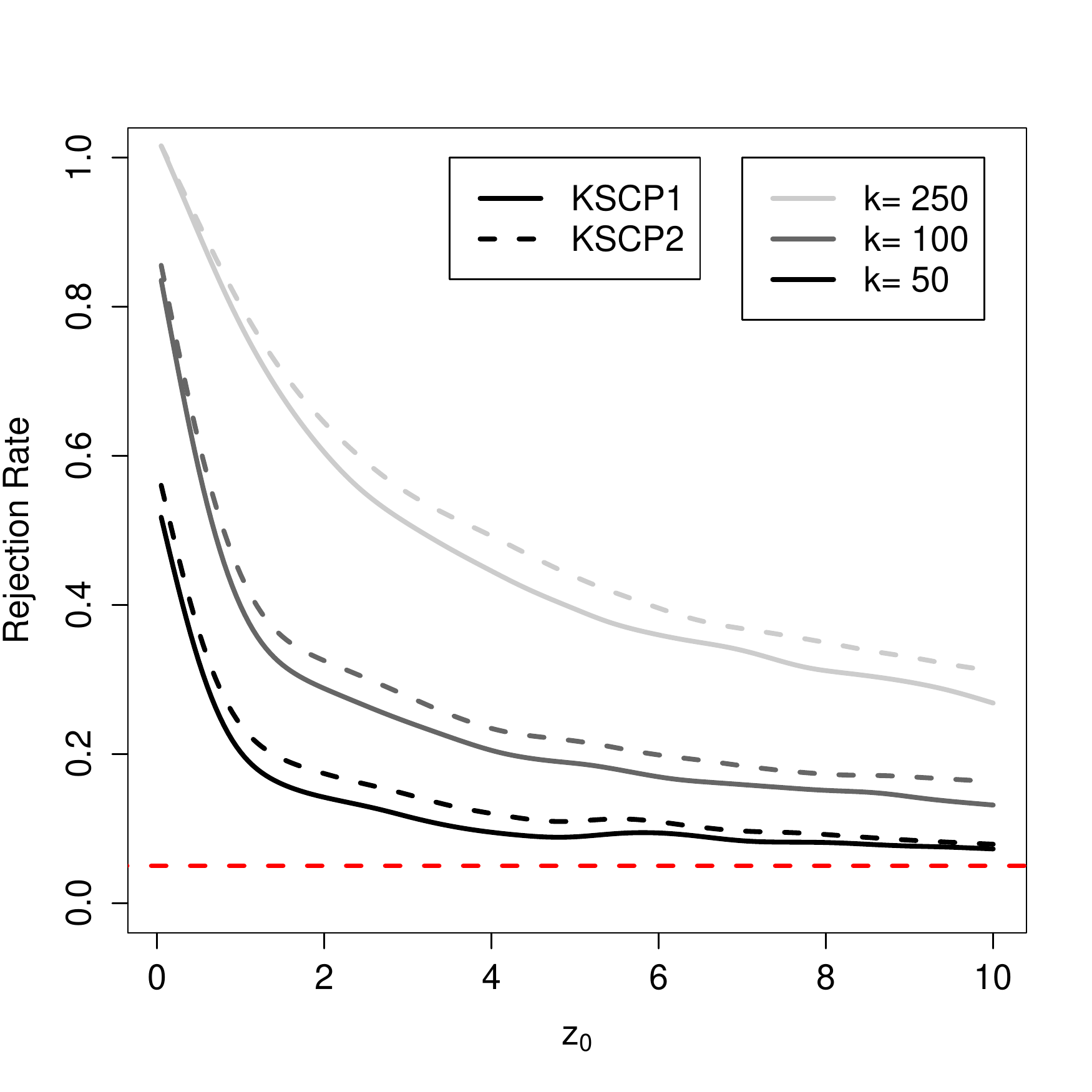}
\includegraphics[width=69mm]{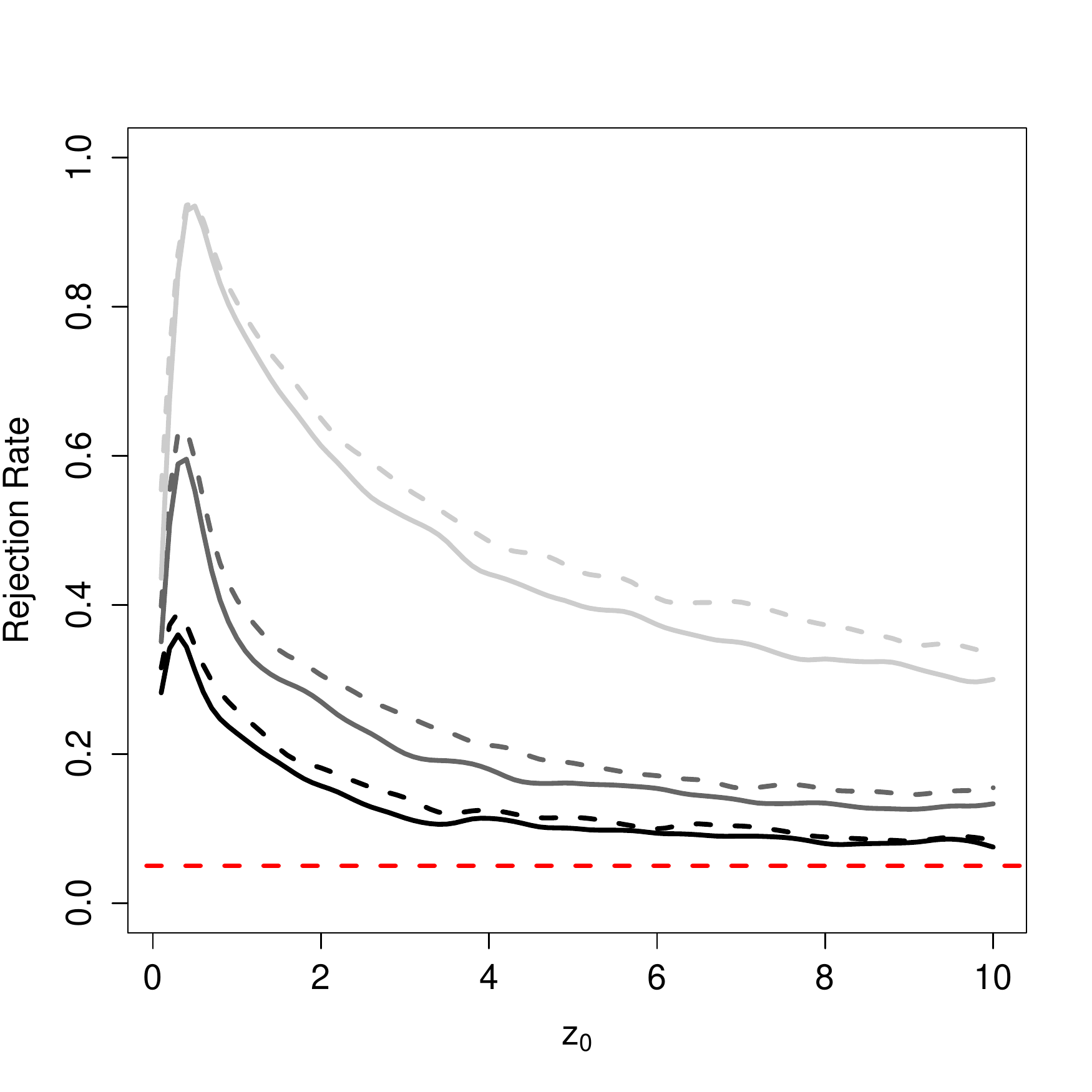}
\vspace{-0.9cm}
\caption{Rejection rates of the KSCP-Test 1 and 2 for different $z_0$. Left panel: pure jump subordinator, right panel: subordinator with a drift plus Brownian motion. }
\label{fig:h1z0}
\end{figure}

\begin{figure}[t!]
\vspace{-.3cm}
\centering
\includegraphics[width=69mm]{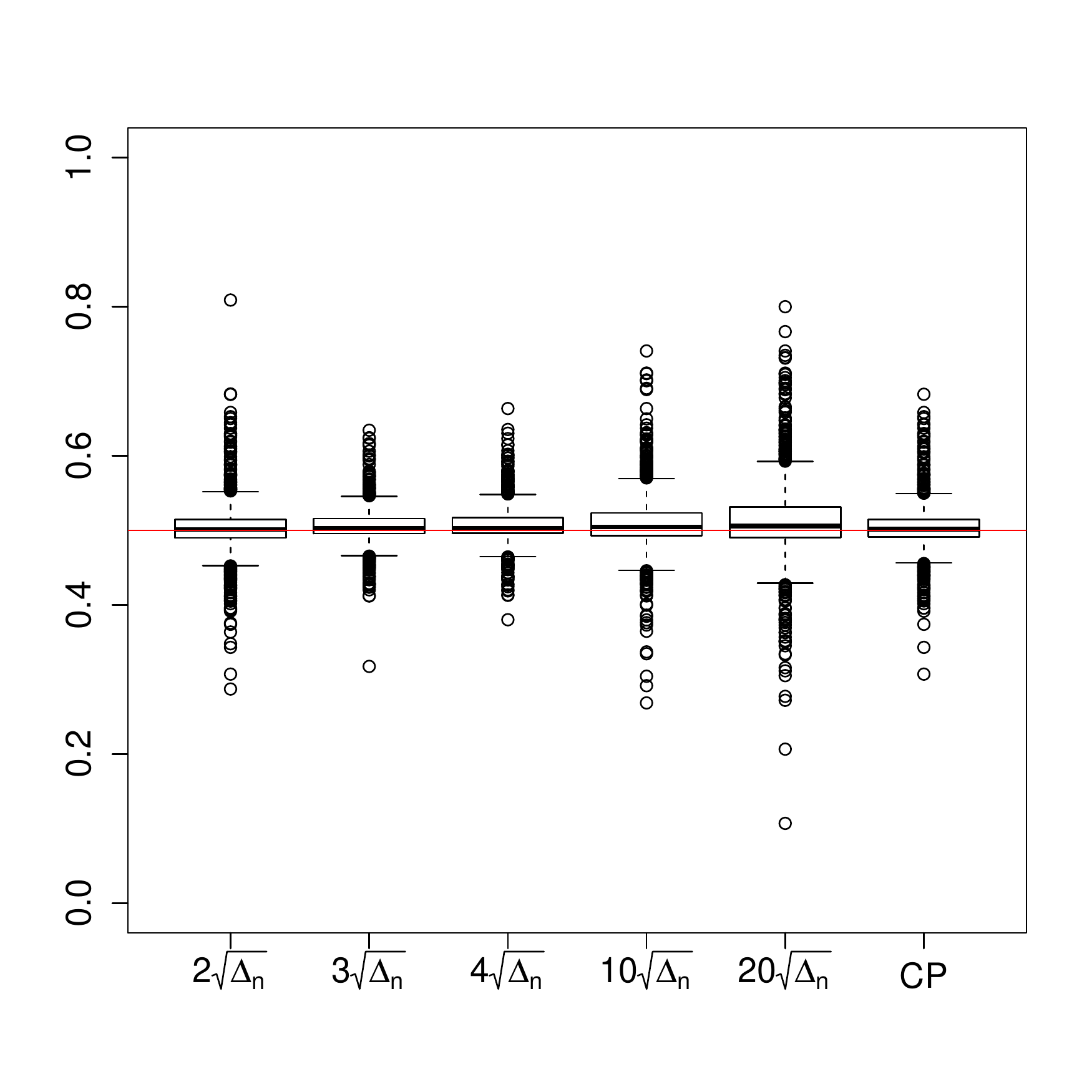}
\includegraphics[width=69mm]{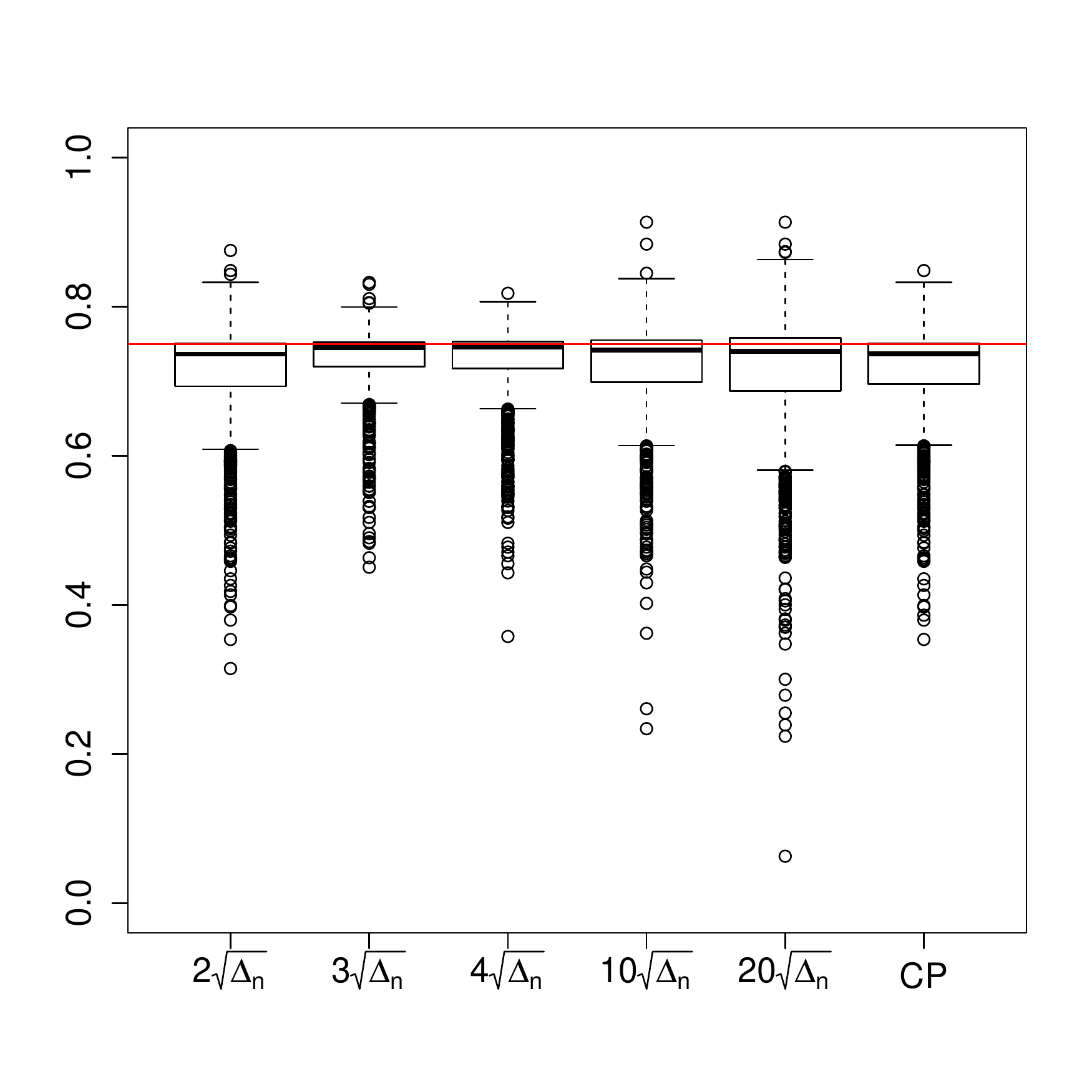}
\vspace{-1.2cm}
\caption{Box plots for the estimators $\tilde \theta_n^{(z_0)}$ and $\hat \theta_n^{(\eps)}$ based on a subordinator with a drift plus Brownian motion and a change from $\beta=1$ to $\beta=4$ at $\theta_0=0.5$ (left panel) and $\theta_0=0.75$ (right panel). The first five box plots in each panel correspond to five different choices of $z_0$.} \label{fig:h1bp}
\end{figure}

The simulation results under these settings are partially reported in Table~\ref{tab:h01} and~\ref{tab:h02} (for the null hypothesis) and in Figures~\ref{fig:h1beta}--\ref{fig:h1bp} (for various alternatives). More precisely, Table~\ref{tab:h01} and~\ref{tab:h02} contain simulated rejection rates under the null hypothesis for various values of $k_n$ and $z_0$ in the KSCP-tests, for the pure jump subordinator (Table~\ref{tab:h01}) and for the process involving a continuous component (Table~\ref{tab:h02}). For the CP-tests, the suprema over $z\in [\eps, \infty)$ were approximated by taking a maximum over a finite grid~$M$: we used the grids $M= \lbrace j \cdot 0.05 \mid j = 1, \ldots, 200 \rbrace$ in the pure jump case, resulting in $\eps=0.05$, and $M= \lbrace (2+ j \cdot 0.5) \sqrt{\Delta_n} \mid j = 0, \ldots, 196 \rbrace$ in the case $b_t = \sigma_t \equiv 1$, resulting in $\eps=2\sqrt \Delta_n$. In the latter case, we chose $\eps$ depending on $\sqrt \Delta_n$ since jumps of smaller size may be dominated by the Brownian component resulting in a loss of efficiency of the CP-test (see also the results in Figure~\ref{fig:h1z0} below). The results in the two tables reveal a rather precise approximation of the nominal level of the tests ($\alpha=5\%$) in all scenarios. In general, KSCP-Test~1 turns out to be slightly more conservative than KSCP-Test~2.

\begin{table}[b]
\begin{center}
\footnotesize{
\begin{tabular}{ c ||c||c|c|c|c|c|c| }
\hline
\multicolumn{1}{|c||}{$k_n$} & \multicolumn{1}{c||}{$\text{CP}$-Test} & \multicolumn{1}{c|}{Pointwise Tests} & 
\multicolumn{1}{c|}{$z_0 = 0.1$} & \multicolumn{1}{c|}{$z_0 = 0.15$} & \multicolumn{1}{c|}{$z_0 = 0.25$} & 
\multicolumn{1}{c|}{$z_0 = 1$} & \multicolumn{1}{c|}{$z_0 = 2$} \\
\hline
\multicolumn{1}{|c||}{$50$} & \multicolumn{1}{c||}{0.06} & \multicolumn{1}{c|}{KSCP-Test 1} & 
\multicolumn{1}{c|}{$0.048$} & \multicolumn{1}{c|}{$0.056$} & \multicolumn{1}{c|}{$0.047$} & 
\multicolumn{1}{c|}{$0.035$} & \multicolumn{1}{c|}{$0.033$} \\
\multicolumn{1}{|c||}{ } & \multicolumn{1}{c||}{ } & \multicolumn{1}{c|}{KSCP-Test 2} & 
\multicolumn{1}{c|}{$0.060$} & \multicolumn{1}{c|}{$0.067$} & \multicolumn{1}{c|}{$0.060$} & 
\multicolumn{1}{c|}{$0.050$} & \multicolumn{1}{c|}{$0.048$} \\
\hline
\multicolumn{1}{|c||}{75} & \multicolumn{1}{c||}{0.054} & \multicolumn{1}{c|}{KSCP-Test 1} & 
\multicolumn{1}{c|}{$0.034$} & \multicolumn{1}{c|}{$0.044$} & \multicolumn{1}{c|}{$0.045$} & 
\multicolumn{1}{c|}{$0.041$} & \multicolumn{1}{c|}{$0.046$} \\
\multicolumn{1}{|c||}{ } & \multicolumn{1}{c||}{ } & \multicolumn{1}{c|}{KSCP-Test 2} & 
\multicolumn{1}{c|}{$0.045$} & \multicolumn{1}{c|}{$0.059$} & \multicolumn{1}{c|}{$0.061$} & 
\multicolumn{1}{c|}{$0.058$} & \multicolumn{1}{c|}{$0.060$} \\
\hline
\multicolumn{1}{|c||}{100} & \multicolumn{1}{c||}{0.06} & \multicolumn{1}{c|}{KSCP-Test 1} & 
\multicolumn{1}{c|}{$0.047$} & \multicolumn{1}{c|}{$0.044$} & \multicolumn{1}{c|}{$0.042$} & 
\multicolumn{1}{c|}{$0.044$} & \multicolumn{1}{c|}{$0.042$} \\
\multicolumn{1}{|c||}{ } & \multicolumn{1}{c||}{ } & \multicolumn{1}{c|}{KSCP-Test 2} & 
\multicolumn{1}{c|}{$0.060$} & \multicolumn{1}{c|}{$0.056$} & \multicolumn{1}{c|}{$0.058$} & 
\multicolumn{1}{c|}{$0.062$} & \multicolumn{1}{c|}{$0.056$} \\
\hline
\multicolumn{1}{|c||}{150} & \multicolumn{1}{c||}{0.06} & \multicolumn{1}{c|}{KSCP-Test 1} & 
\multicolumn{1}{c|}{$0.049$} & \multicolumn{1}{c|}{$0.056$} & \multicolumn{1}{c|}{$0.049$} & 
\multicolumn{1}{c|}{$0.040$} & \multicolumn{1}{c|}{$0.042$} \\
\multicolumn{1}{|c||}{ } & \multicolumn{1}{c||}{ } & \multicolumn{1}{c|}{KSCP-Test 2} & 
\multicolumn{1}{c|}{$0.065$} & \multicolumn{1}{c|}{$0.064$} & \multicolumn{1}{c|}{$0.065$} & 
\multicolumn{1}{c|}{$0.059$} & \multicolumn{1}{c|}{$0.061$} \\
\hline
\multicolumn{1}{|c||}{250} & \multicolumn{1}{c||}{0.07} & \multicolumn{1}{c|}{KSCP-Test 1} & 
\multicolumn{1}{c|}{$0.046$} & \multicolumn{1}{c|}{$0.042$} & \multicolumn{1}{c|}{$0.046$} & 
\multicolumn{1}{c|}{$0.055$} & \multicolumn{1}{c|}{$0.050$} \\
\multicolumn{1}{|c||}{ } & \multicolumn{1}{c||}{ } & \multicolumn{1}{c|}{KSCP-Test 2} & 
\multicolumn{1}{c|}{$0.054$} & \multicolumn{1}{c|}{$0.048$} & \multicolumn{1}{c|}{$0.059$} & 
\multicolumn{1}{c|}{$0.072$} & \multicolumn{1}{c|}{$0.060$} \\
\hline
\end{tabular}
}
\caption{\textbf{Test procedures under} $\textbf H_0$. 
          Simulated relative frequency of rejections in the application of the KSCP-Test 1, the KSCP-Test 2 and the 
					$\text{CP}$-Test to $1000$ pure jump subordinator data vectors under the null hypothesis.} 
\label{tab:h01}
\end{center}
\vspace{-.1cm}
\end{table}

\begin{table}[t]
\begin{center}
\footnotesize{
\begin{tabular}{ c ||c||c|c|c|c|c|c|| }
\hline
\multicolumn{1}{|c||}{$k_n$} & \multicolumn{1}{c||}{${\text{CP}}$-Test} & \multicolumn{1}{c|}{Pointwise Tests} & 
\multicolumn{1}{c|}{$z_0 = 2 \sqrt{\Delta_n}$} & \multicolumn{1}{c|}{$z_0 = 3.5 \sqrt{\Delta_n}$} & \multicolumn{1}{c|}{$z_0 = 6.5 \sqrt{\Delta_n}$} & \multicolumn{1}{c|}{$z_0 = 7 \sqrt{\Delta_n}$} \\
\hline
\multicolumn{1}{|c||}{$50$} & \multicolumn{1}{c||}{0.049} & \multicolumn{1}{c|}{KSCP-Test 1} & 
\multicolumn{1}{c|}{$0.032$} & \multicolumn{1}{c|}{$0.036$} &  
\multicolumn{1}{c|}{$0.035$} & \multicolumn{1}{c|}{$0.031$} \\
\multicolumn{1}{|c||}{ } & \multicolumn{1}{c||}{ } & \multicolumn{1}{c|}{KSCP-Test 2} & 
\multicolumn{1}{c|}{$0.049$} & \multicolumn{1}{c|}{$0.051$} &  
\multicolumn{1}{c|}{$0.049$} & \multicolumn{1}{c|}{$0.050$} \\
\hline
\multicolumn{1}{|c||}{75} & \multicolumn{1}{c||}{0.050} & \multicolumn{1}{c|}{KSCP-Test 1} & 
\multicolumn{1}{c|}{$0.042$} & \multicolumn{1}{c|}{$0.039$} &  
\multicolumn{1}{c|}{$0.039$} & \multicolumn{1}{c|}{$0.032$} \\
\multicolumn{1}{|c||}{ } & \multicolumn{1}{c||}{ } & \multicolumn{1}{c|}{KSCP-Test 2} & 
\multicolumn{1}{c|}{$0.050$} & \multicolumn{1}{c|}{$0.057$} & 
\multicolumn{1}{c|}{$0.051$} & \multicolumn{1}{c|}{$0.053$} \\
\hline
\multicolumn{1}{|c||}{100} & \multicolumn{1}{c||}{0.051} & \multicolumn{1}{c|}{KSCP-Test 1} & 
\multicolumn{1}{c|}{$0.039$} & \multicolumn{1}{c|}{$0.040$} &  
\multicolumn{1}{c|}{$0.037$} & \multicolumn{1}{c|}{$0.038$} \\
\multicolumn{1}{|c||}{ } & \multicolumn{1}{c||}{ } & \multicolumn{1}{c|}{KSCP-Test 2} & 
\multicolumn{1}{c|}{$0.051$} & \multicolumn{1}{c|}{$0.054$} &  
\multicolumn{1}{c|}{$0.049$} & \multicolumn{1}{c|}{$0.057$} \\
\hline
\multicolumn{1}{|c||}{150} & \multicolumn{1}{c||}{0.057} & \multicolumn{1}{c|}{KSCP-Test 1} & 
\multicolumn{1}{c|}{$0.038$} & \multicolumn{1}{c|}{$0.045$} &  
\multicolumn{1}{c|}{$0.034$} & \multicolumn{1}{c|}{$0.039$} \\
\multicolumn{1}{|c||}{ } & \multicolumn{1}{c||}{ } & \multicolumn{1}{c|}{KSCP-Test 2} & 
\multicolumn{1}{c|}{$0.057$} & \multicolumn{1}{c|}{$0.057$} &  
\multicolumn{1}{c|}{$0.053$} & \multicolumn{1}{c|}{$0.052$} \\
\hline
\multicolumn{1}{|c||}{250} & \multicolumn{1}{c||}{0.049} & \multicolumn{1}{c|}{KSCP-Test 1} & 
\multicolumn{1}{c|}{$0.031$} & \multicolumn{1}{c|}{$0.035$} &  
\multicolumn{1}{c|}{$0.042$} & \multicolumn{1}{c|}{$0.030$} \\
\multicolumn{1}{|c||}{ } & \multicolumn{1}{c||}{ } & \multicolumn{1}{c|}{KSCP-Test 2} & 
\multicolumn{1}{c|}{$0.049$} & \multicolumn{1}{c|}{$0.048$} &  
\multicolumn{1}{c|}{$0.053$} & \multicolumn{1}{c|}{$0.042$} \\
\hline
\end{tabular}
}
\caption{\textbf{Test procedures under} $\textbf H_0$. 
          Simulated relative frequency of rejections in the application of the KSCP-Test 1, the KSCP-Test 2 and the 
					$\text{CP}$-Test to $1000$ subordinator data vectors plus a drift $b =1$ and plus a Brownian motion 
					under $\textbf H_0$.}
\label{tab:h02}
\end{center}
\vspace{-.4cm}
\end{table}

The results presented in Figure~\ref{fig:h1beta} consider the CP-test for alternatives involving one fixed break point at $\theta_0=0.5$ and a varying height of the jump size, as measured through the value of $\beta$ in \eqref{eq:ubeta}.  In contrast to the results in Tables~\ref{tab:h01} and \ref{tab:h02}, due to computational reasons, we subsequently used smaller grids $M= \lbrace j \cdot 0.2 \mid j=1, \ldots ,20 \rbrace$ for the case $b_t =\sigma_t \equiv 0$, resulting in $\eps=0.2$, and $M= \lbrace 2.5 \cdot \sqrt{\Delta_n} \cdot j \mid j = 1, \ldots , 20 \rbrace$ for the case $b_t =\sigma_t \equiv 1$, resulting in $\eps=2.5 \sqrt \Delta_n$. The left plot is based on the pure jump process ($b_t = \sigma_t \equiv 0$), whereas the right one is based on $b_t = \sigma_t \equiv 1$. The dashed red line indicates the nominal level of $\alpha=5\%$. We observe that the rejection rate of the test is increasing in $\beta$ (as to be expected) and in $k_n$. The latter can be explained by the fact that $k_n$ represents the effective sample size (interpretable as the number of trading days). Finally, the rejection rates turn out to be higher when no continuous component is involved in the underlying semimartingale.

The next two graphics in Figure~\ref{fig:h1theta} show the rate of rejection of the CP-Test under alternatives involving one break point from $\beta=1$ to $\beta=2.5$ within the model in \eqref{eq:ubeta} for varying locations of the change point $\theta_0\in(0,1)$. Again, the left and right plots correspond to $b_t=\sigma_t\equiv0$ and $\equiv 1$, respectively. 
Additionally to the general conclusions drawn from the results in Figure~\ref{fig:h1beta}, we observe that break points can be detected best if $\theta_0=1/2$, and that the rejection rates are symmetric around that point.


Figure~\ref{fig:h1z0} shows the rejection rates of the KSCP-Test 1 and 2, evaluated at different points $z_0$, for one fixed alternative model involving a single change from $\beta =1$ to $\beta =2.5$ at the point $\theta_0 = 1/2$. The curves in the left plot are based on a pure jump process. We can see that the rejection rates are decreasing in $z_0$, explainable by the fact that there are only very few large jumps both for $\beta =1$ and for $\beta =2.5$. In the right plot, involving drift and volatility ($b_t=\sigma_t \equiv 1$), we observe a maximal value of the rejection rates that is increasing in the number of trading days, $k_n$. 
For values of $z_0$ smaller than this maximum, the contribution of the Brownian component (an independent normally distributed term with variance $\Delta_n$ within each increment $\Delta_j^n X$) predominates the jumps of that size and results in a decrease of the rejection rate.

Finally, in Figure \ref{fig:h1bp}, we depict box plots for the estimators $\tilde \theta_n^{\scriptscriptstyle (z_0)}$ and $\hat \theta_n^{\scriptscriptstyle (\eps)}$ of the change point for certain values of $z_0$ and for $M$ as specified in the case of Tables~\ref{tab:h01} and \ref{tab:h02}. The results are based on two models, involving a change in $\beta$ from $1$ to $4$ at time point $\theta_0=0.5$ (left panel) and $\theta_0=0.75$ (right panel) for $k_n = 250$ and $\Delta_n^{-1} = 90$, and with $b_t=\sigma_t\equiv 1$. We observe a reasonable approximation of the true value (indicated by the red line) with more accurate approximations for $\theta_0=0.5$. For $\theta_0=0.75$, the distribution of the estimator is skewed, giving more weight to the left tail directing to $\theta_0=0.5$. This might be explained by the fact that the distribution of the argmax absolute value of a tight-down stochastic process indexed by $\theta\in [0,1]$ gives very small weight to the boundaries of the unit interval.
Moreover, as for the results presented in the right plot of Figure~\ref{fig:h1z0}, the plots in Figure \ref{fig:h1bp} reveal that the estimator $\tilde \theta_n^{\scriptscriptstyle (z_0)}$ behaves best for an intermediate choice of $z_0$. Results for $b_t=\sigma_t\equiv0$ are not depicted for the sake of brevity, since they do not transfer any additional insight.


\section{Appendix} \label{sec:proof}
\def\theequation{6.\arabic{equation}}
\setcounter{equation}{0}

\subsection{Proof of Lemma \ref{WahrschAbschLem}}

Let $\eps < (\delta/6 \wedge 1)$ and pick a smooth cut-off function $c_\eps: \mathbb R \to \mathbb R$ satisfying
\begin{align*}
1_{[-\eps/2,\eps/2]}(u) \leq c_\eps(u) \leq 1_{[-\eps,\eps]}(u).
\end{align*}
We also define the function $\bar c_\eps$ via $\bar c_\eps(u)=1-{c_\eps}(u)$. We use $\bar c_{\epsilon}$ to define the ``large'' jumps of the process, that means, there exist independent processes $X^\eps$ and $\widetilde{X}^\eps$ such that $X =_d X^\eps + \widetilde{X}^\eps$ where $\widetilde{X}^\eps$ is a compound Poisson process with intensity $\lambda_{\eps}=\int \bar c_\eps(u) \nu(du)$ and jump distribution $\rho_\eps(du)=\bar c_\eps(u) \nu(du)/\lambda_{\eps}$. See e.g.\ \cite{FigLopHou09}. Accordingly, $X^\eps$ is an It\=o semimartingale with characteristics $(b^\eps_s,\sigma^2_s,c_\eps(u)\nu(du))$, where we set $b_s^{\epsilon}=b_s-\int 1_{\{|u| > 1\}} u \bar c_\eps(u) \nu(du)$.

Since our result is a distributional one only, it is possible to work with this particular representation of $X$ in the following. Call $N_t^\eps$ the number of jumps of $\widetilde{X}^\eps$ up to time $t$. Define $f(x)=1_{\{x \ge z \}}$.
Using the law of total expectation we have
\begin{multline} \label{eq:fdec}
\Eb[f(X_t)] = \sum_{k=0}^{\infty} e^{-\lambda_\eps t} \frac{(\lambda_\eps t)^k}{k!} \Eb[f(X_t)|N_t^\eps = k] \\ = e^{-\lambda_\eps t} \Eb[f(X^\eps_t)]  + e^{-\lambda_\eps t} \lambda_\eps t \Eb[f(X^\eps_t + \xi_1)]  \\ + \sum_{k=2}^{\infty} e^{-\lambda_\eps t} \frac{(\lambda_\eps t)^k}{k!} \Eb\Big[f\Big(X^\eps_t+\sum_{\ell=1}^k \xi_\ell \Big) \Big],
\end{multline}
where the random variables $\xi_\ell$ are i.i.d.\ with distribution $\rho_\eps$.

For the first summand on the right of the last display, i.e.\ the case of no large jumps, we discuss drift, volatility and small jumps separately. For that purpose,  we write
$
X^\eps_t = B^{\eps}_t + Q_t + Y_t^{\eps}
$
where
$B^{\eps}_t = \int_0^t b_s^\epsilon ds$,  $Q_t = \int_0^t \sigma_s dW_s$
and where $Y_t^{\eps}$ is a pure jump L\'evy martingale with jump measure $c_\eps(u)\nu(du)$. By the triangle inequality
\[
e^{-\lambda_\eps t} \Eb[f(X^\eps_t)] \leq P(X^\eps_t \geq \delta) \leq P(B^\eps_t \geq \delta/3) + P(C_t \geq \delta/3) + P(Y^\eps_t \geq \delta/3).
\]
Let us show that the right-hand side of this display can be bounded by $Kt^2$ for all $0<t<t_0$, with constants $K=K(\delta)$ and $t_0=t_0(\delta)$.
Regarding the summand $P(Y^\eps_t \geq \delta/3)$, we can use equation (3.3) in \cite{FigLopHou09} applied to a pure jump L\'evy process. Following their result, $\eps < \delta/6$ ensures the existence of $K$ and $t_0$, both depending on $\delta$ only, such that
\begin{align} \label{SmaJum}
\Prob(Y_t^{\eps} \geq \delta/3) < K t^2
\end{align}
for all $0 < t < t_0$. Since $b$ and $\sigma$ are bounded, we further have
$
\Eb[|B^\eps_t|^r] \leq K t^r$ and
$
\Eb[|Q_t|^r] \leq K t^{r/2}
$
for an arbitrary integer $r$ and with $K$ depending on $\delta$ again. Markov's inequality then yields bounds similar to (\ref{SmaJum}) when applied to the processes involving drift, $B_t^\eps$, and volatility, $C_t$.

Also, for the sum over $k$ on the right-hand side of \eqref{eq:fdec}, we have
\begin{align*}
e^{-\lambda_\eps t} \sum_{k=2}^{\infty} \frac{(\lambda_\eps t)^k}{k!} < Kt^2.
\end{align*}

It therefore remains to focus on $\Eb[f(X^\eps_t + \xi_1)]$. As a consequence of Con\-di\-tion~\ref{Assump1}(d), and observing that the distribution of $\xi_1$ is $h(u) \bar c_\eps(u) du/\lambda_\eps$ with the L\'evy density $h$,  it follows that
\begin{align*}
g(x) = \Eb[f(x+\xi_1)] = \Prob(x + \xi_1 \ge z)
\end{align*}
is twice continuously differentiable with bounded derivatives.  Using independence of $X^\eps$ and $\xi_1$, it is sufficient to discuss $\Eb[g(X_t^\eps)]$, for which we can use It\^o formula now: for arbitrary $Y$ we have
\begin{multline} \label{decomp} 
	g(Y_t) = g( Y_0) + \int_{0}^t g'(Y_{s-}) dY_s + \frac 12 \int_{0}^t g''(Y_{s-}) d[Y, Y]^c_s \\
	+ \sum_{0<s \leq t} \left(g(Y_s) - g(Y_{s-}) - g'(Y_{s-}) \Delta Y_s) \right),
 \end{multline}
where $[Y, Y]^c_s$ denotes the quadratic variation and $\Delta Y_s$ is the jump size at time $s$. Plugging in $X^\eps$ for $Y$ we discuss each of the four summands in \eqref{decomp} separately: 
first, $u \ge z$ implies $\bar c_{\eps}(u) = 1$ by definition of $\eps$. Thus, with $X_0^\eps = 0$
\begin{align*}
g(X_0^\eps) = \Prob(\xi_1 \ge z )
=
\frac{1}{\lambda_\eps} \int 1_{\{ u \ge z\}} h(u) \bar c_{\eps}(u) du
=
\frac{1}{\lambda_\eps} \int 1_{\{ u \ge z\}} h(u) du
=
\frac{1}{\lambda_\eps} \nu([z, \infty)).
\end{align*}
Second, two of the three summands in $X^\eps_s$ are martingales. Therefore
\begin{align*}
\left \vert \Eb\left[\int_{0}^t g'(X^\eps_{s-}) dX^\eps_s\right] \right \vert \leq \int_{0}^t \left \vert \Eb[g'(X^\eps_{s-})] b^\eps_s \right \vert ds < Kt
\end{align*}
due to boundedness of the first derivatives of $g$. We may proceed similarly for the third term in (\ref{decomp}). Finally, conditioning on $X^\eps_{s-}$ and using the definition of a compensator gives
\begin{multline*}
\sum_{0 < s \leq t} \Eb\left[g(X^\eps_s) - g(X^\eps_{s-}) - g'(X^\eps_{s-}) \Delta X^\eps_s \right] \\
=
\int_0^t \int \Eb[g(X_{s-}^\eps +u) - g(X_{s-}^\eps) - g'(X_{s-}^\eps) u] c_\eps(u) \nu(du) ds
\end{multline*}
for the final quantity. The Taylor formula proves that the inner integrand above may be bounded in absolute value by $K u^2  c_\eps(u)$. Since $\nu$ is a L\'evy measure, we obtain
\begin{align*}
\big \vert \Eb[g(X_t^\eps)]-\frac{1}{\lambda_\eps} \nu([z,\infty))\big \vert < Kt.
\end{align*}
From $|1-\exp(-\lambda_\eps t)| < Kt$ for $0<t<t_0$ the conclusion follows. \qed \\

\subsection{Proof of Theorem \ref{WeakConvThm}}
Due to Theorem 1.6.1 in \cite{VanWel96}, it suffices to prove weak convergence $\Gb_n \weak \Gb$ in  $\ell^{\infty}(A_k)$ for any fixed $k\in \N$.

To this end, we   use Theorem 11.16 of \cite{Kos08}. Note that $\Gb_n$ can be written as
\[
\Gb_n(\theta,z)
=
\frac{1}{\sqrt{k_n}} \sum \limits_{j=1}^{\lfloor n\theta \rfloor} \{ \mathtt 1_{\lbrace \Delta_j^n X \geq z \rbrace} - \mathbb P (\Delta_j^n X \geq z) \}
=
\sum \limits_{j = 1}^n  \{  f_{nj} (\omega ; \theta,z) - \mathbb E f_{nj} ( \cdot ; \theta,z) \}
\]
with the triangular array $\lbrace f_{nj}(\omega ; \theta,z) \mid n \geq 1; j=1, \ldots ,n; (\theta,z) \in A_k \rbrace$ consisting of the processes
$$f_{nj}(\theta,z) \defeq  f_{nj}(\omega ;\theta,z)  \defeq \frac{1}{\sqrt{k_n}} \mathtt 1_{\lbrace j \leq \lfloor n\theta \rfloor \rbrace}
\mathtt 1_{\lbrace \Delta_j^n X(\omega) \geq z \rbrace},$$
which are independent within rows since we assume a deterministic drift and volatility. By Theorem 11.16 in \cite{Kos08}, the proof is complete if   the following six conditions for $\lbrace f_{nj} \rbrace$ can be established:
\begin{enumerate}
\item[(1)]  $\lbrace f_{nj} \rbrace$ is almost measurable Suslin (AMS);
\item[(2)] the $\lbrace f_{nj} \rbrace$ are manageable with envelopes $\lbrace F_{nj} \mid n \in \mathbb N, j= 1, \ldots,n \rbrace$ given
      through $F_{nj} \defeq k_n^{-1/2} \mathtt 1_{\lbrace \Delta_j^n X \geq k^{-1} \rbrace}$, which are also independent within rows;
\item[(3)]  $H(\theta_1, z_1 ; \theta_2, z_2) = \lim \limits_{n \rightarrow \infty} \mathbb E \{ \Gb_n (\theta_1, z_1) \Gb_n (\theta_2 ,z_2) \}$ for all
      $(\theta_1, z_1) , (\theta_2, z_2) \in A_k$;
\item[(4)]  $\limsup \limits_{n \rightarrow \infty} \sum \limits_{j =1}^n \mathbb E^{\ast} F_{nj}^2 < \infty$;
\item[(5)]  $\lim \limits_{n \rightarrow \infty} \sum \limits_{j =1}^n \mathbb E^{\ast} F_{nj}^2 \mathtt 1_{\lbrace F_{nj} > \eps \rbrace}=0$
      for all $\eps >0$;
\item[(6)]  $\rho(\theta_1, z_1; \theta_2, z_2) = \lim \limits_{n \rightarrow \infty} \rho_n(\theta_1,z_1;\theta_2,z_2)$ for every
      $(\theta_1,z_1),(\theta_2,z_2) \in A_k$, where
	\[
	\rho_n(\theta_1,z_1;\theta_2,z_2) \defeq \bigg\{ \sum \limits_{j=1}^n \mathbb E \left| f_{nj} (\cdot; \theta_1,z_1) -
			f_{nj}( \cdot; \theta_2,z_2) \right|^2 \bigg\}^{\frac{1}{2}}.
	\]
	Moreover, $\rho_n(\theta_1^{(n)}, z_1^{(n)} ; \theta_2^{(n)}, z_2^{(n)}) \rightarrow 0$ for all sequences
			$(\theta_1^{(n)},z_1^{(n)})_{n \in \mathbb N} $ and $(\theta_2^{(n)},z_2^{(n)})_{n \in \mathbb N} \subset A_k$ such that
			$\rho(\theta_1^{(n)}, z_1^{(n)} ; \theta_2^{(n)}, z_2^{(n)}) \rightarrow 0$.
\end{enumerate}

\smallskip
\noindent
\textit{Proof of {\rm (1)}.}
By Lemma 11.15 in \cite{Kos08}, the triangular array $\lbrace f_{nj} \rbrace$ is AMS provided it is separable, i.e., provided
for every $n \in \mathbb N$, there exists a countable subset $S_n \subset A_k$, such that
$$
\mathbb P^{\ast}\bigg(\sup \limits_{(\theta_1,z_1) \in A_k} \inf \limits_{(\theta_2,z_2) \in S_n} \sum \limits_{j=1}^n  \{ f_{nj}(\omega; \theta_2,z_2) -
f_{nj}(\omega; \theta_1,z_1) \} ^2 >0\bigg)=0.
$$
Define $S_n \defeq \mathbb Q^2 \cap A_k$ for all $n \in \mathbb N$.
Then, for every element $\omega$ of the underlying probability space and for every $(\theta_1,z_1) \in A_k$, there exists an
$(\theta_2,z_2) \in S_n$ such that
\[
\sum \limits_{j=1}^n \{ f_{nj}(\omega; \theta_2,z_2) - f_{nj}(\omega; \theta_1,z_1) \}^2 =0.
\]

\smallskip
\noindent
\textit{Proof of {\rm (2)}.}
The $\lbrace F_{nj} \rbrace$ are independent within rows since we assume deterministic characteristics of the underlying process. Therefore, according to Theorem 11.17 in \cite{Kos08}, it suffices to prove that the triangular arrays
$$\lbrace \tilde f_{nj} (\omega;z) \defeq k_n^{-1/2} \mathtt 1_{\lbrace \Delta_j^n X \geq z \rbrace} \mid n \in \mathbb N; j=1, \ldots, n; z \in [k^{-1}, \infty) \rbrace,$$
and
$$\lbrace \tilde g_{nj} (\omega;\theta) \defeq \mathtt 1_{\lbrace j \leq \lfloor n\theta \rfloor \rbrace} \mid n \in \mathbb N; j=1, \ldots, n; \theta \in [0,1]\rbrace$$
are manageable with envelopes
$\lbrace \tilde F_{nj} (\omega) \defeq k_n^{-1/2} \mathtt 1_{\lbrace \Delta_j^n X \geq k^{-1} \rbrace} \mid n \in \mathbb N; j=1, \ldots,n \rbrace$ and
$\lbrace \tilde G_{nj} (\omega) :\equiv 1 \mid n \in \mathbb N; j=1, \ldots,n \rbrace$, respectively.

Concerning the first triangular array $\lbrace \tilde f_{nj} \rbrace$ define, for $\omega \in \Omega$ and $n \in \mathbb N$,
$$\mathcal F_{n \omega} \defeq \big\{ ( k_n^{-1/2}\mathtt 1_{\lbrace \Delta_1^n X(\omega) \geq z \rbrace} , \ldots , k_n^{-1/2} \mathtt 1_{\lbrace \Delta_n^n X(\omega) \geq z \rbrace}) \mid z \in [k^{-1}, \infty) \big\} \subset \R^n.$$
For any $j_1,j_2 \in \lbrace 1, \ldots,n \rbrace$, the projection $p_{j_1,j_2}(\mathcal F_{n \omega})$ of $\mathcal F_{n \omega}$ onto the
$j_1$-th and the $j_2$-th coordinate is an element of the set
\begin{multline*}
\Big\{ \lbrace (0,0) \rbrace,
\lbrace (0,0), (k_n^{-1/2},0) \rbrace,
\lbrace (0,0), (0,k_n^{-1/2}) \rbrace,
\lbrace (0,0), (k_n^{-1/2}, k_n^{-1/2} ) \rbrace, \\
\lbrace (0,0), (k_n^{-1/2},0), (k_n^{-1/2}k_n^{-1/2}) \rbrace,
\lbrace (0,0), (0,k_n^{-1/2}), (k_n^{-1/2},k_n^{-1/2}) \rbrace \Big\}.
\end{multline*}
Hence, for every $t \in \mathbb R^2$, no proper coordinate projection of $\mathcal F_{n \omega}$ can surround $t$ in the sense of Definition~4.2 of \cite{Pol90}. Thus, $\mathcal F_{n \omega}$ is a subset of $\mathbb R^n$ of pseudodimension at most $1$ (Definition 4.3 in
\citealp{Pol90}). Additionally, $\mathcal F_{n \omega}$ is a bounded set, whence Corollary 4.10 in \cite{Pol90} yields the existence of constants $A$ and $W$, depending only on the pseudodimension, such that
\[
D_2( x \| \alpha \odot \tilde F_n (\omega) \|_2, \alpha \odot \mathcal F_{n \omega}) \leq A x^{-W} =: \lambda(x),
\]
for all $0<x \leq 1$, for every rescaling vector $\alpha \in \mathbb R^n$ with non-negative entries and for all $\omega \in \Omega$ and
$n \in \mathbb N$. Therein, $\| \cdot \|_2$ denotes the Euclidean distance, $D_2$ denotes the packing number with respect to the Euclidean distance and $\tilde F_n (\omega) \defeq (\tilde F_{n1} (\omega), \ldots , \tilde F_{nn} (\omega)) \in \mathbb R^n$ is the vector of envelopes.
Since $\int_0^1 \sqrt{\log \lambda(x)} dx < \infty$, the triangular array
$\lbrace \tilde f_{nj} \rbrace$ is indeed manageable with envelopes $\lbrace \tilde F_{nj} \rbrace$.

Concerning the triangular array $\lbrace \tilde g_{ni} \rbrace$, we proceed similar and consider  the set
\begin{multline*}
\mathcal G_{n \omega}
\defeq
\lbrace (\tilde g_{n1}(\omega;\theta), \ldots , \tilde g_{nn}(\omega;\theta)) \mid \theta \in [0,1] \rbrace \\
=
\lbrace (0, \ldots , 0) , (1,0, \ldots ,0),(1,1,0, \ldots,0), \ldots, (1, \ldots ,1) \rbrace.
\end{multline*}
Then, for any $j_1,j_2 \in \lbrace 1, \ldots,n \rbrace$, the projection $p_{j_1,j_2}(\mathcal G_{n \omega})$ of $\mathcal G_{n \omega}$ onto the
$j_1$-th and the $j_2$-th coordinate is either $\lbrace (0,0), (1,0) , (1,1) \rbrace$ or $\lbrace (0,0), (0,1) , (1,1) \rbrace$. Therefore, the same reasoning as above shows that $\mathcal G_{n \omega}$ is a set of pseudodimension at most one, whence the triangular
array $ \lbrace \tilde g_{nj} \rbrace$ is manageable with envelopes $\lbrace \tilde G_{nj} \rbrace$.

\smallskip
\noindent
\textit{Proof of {\rm (3)}.}
For any $(\theta_1, z_1), (\theta_2, z_2) \in A_k$, by independence of $\lbrace f_{nj} \rbrace$  within rows, we can write
\begin{align}
\label{CovGnEqn}
& \hspace{-.7cm}
\Exp\{ \Gb_n(\theta_1,z_1) \Gb_n(\theta_2,z_2)  \}  \nonumber \\
&=
 \sum_{j=1}^n  \Exp\big[  \{f_{nj}(\omega; \theta_1,z_1) - \Exp f_{nj}(\cdot; \theta_1,z_1) \}  \{ f_{nj}(\omega; \theta_2,z_2)- \Exp f_{nj}(\cdot; \theta_2,z_2) \}\big] \nonumber \\
&=
\frac{1}{k_n} \sum_{j=1}^{\ip{n(\theta_1 \wedge \theta_2)}} \{ \Prob(\Delta_j^n X \geq z_1 \vee z_2) - \Prob(\Delta_j^n X \geq z_1)\Prob(\Delta_j^n X \geq z_2) \}
\end{align}
By Remark~\ref{WahrschAbschRem} and the choice $K=K(k^{-1})$ and $t_0=t_0(k^{-1}) >0$, we have
\begin{align}
\label{Wahrschabschforallieqn}
\Prob(\Delta_j^n X \geq z)=\Delta_n \nu([z, \infty)) + O(\Delta_n^2), \quad n\to\infty
\end{align}
for all $z \geq k^{-1}$ and all $j = 1, \ldots ,n$, whence the right-hand side of equation \eqref{CovGnEqn} can be written as
\[
\frac{\ip{n(\theta_1 \wedge \theta_2)}}{n} \{ \nu([z_1 \vee z_2, \infty))  + O(\Delta_n)  \} = H(\theta_1, z_1;\theta_2,z_2) + o(1), \quad n \to \infty.
\]

\smallskip
\noindent
\textit{Proof of {\rm (4)}.}
Again from Remark~\ref{WahrschAbschRem}, we have
\begin{align*}
\sum \limits_{j=1}^n \Exp^{\ast} F_{nj}^2
=
\frac{1}{n \Delta_n} \sum \limits_{j=1}^n \mathbb P(\Delta_j^n X \geq k^{-1})
=
\nu([k^{-1}, \infty)) +O(\Delta_n)
\rightarrow
\nu([k^{-1}, \infty)) < \infty
\end{align*}
as $n\to \infty$.

\smallskip
\noindent
\textit{Proof of {\rm (5)}.}
For $\eps >0$ define $N \defeq \min \lbrace n \in \mathbb N \mid k_m^{-1/2} \leq \eps \text{ for all }
m \geq n \rbrace$. Choose $K=K(k^{-1})$ and $t_0 = t_0(k^{-1})$ as in Lemma~\ref{WahrschAbschLem}. Then, for any sufficiently large $n$ such that  $\Delta_n < t_0$, we have
\begin{multline*}
\sum \limits_{j=1}^n \mathbb E^{\ast} F_{nj}^2 \mathtt 1_{\lbrace F_{nj} > \eps \rbrace}
\leq
\sum \limits_{j=1}^N \mathbb E^{\ast} F_{nj}^2
=
\frac{1}{n \Delta_n} \sum \limits_{j=1}^N \mathbb P( \Delta_j^n X \geq k^{-1}) \\
\leq
\frac{N}{n}  \{ \nu([k^{-1}, \infty)) + K \Delta_n \} \to 0, \qquad n \to \infty.
\end{multline*}

\smallskip
\noindent
\textit{Proof of {\rm (6)}.}
For $(\theta_1,z_1),(\theta_2,z_2) \in A_k$, we can write
\begin{align*}
\rho_n^2 (&\theta_1,z_1;\theta_2,z_2) \\
&=
\sum \limits_{j=1}^n \mathbb E \left| f_{nj}( \cdot ; \theta_1,z_1) - f_{nj}( \cdot; \theta_2,z_2) \right|^2\\
&=
\frac{1}{n \Delta_n} \bigg\{ \sum_{j=1}^{\lfloor n (\theta_1 \wedge \theta_2) \rfloor} (\mathbb P(\Delta_j^n X \geq z_1 \wedge z_2) - \mathbb P( \Delta_j^n X \geq z_1 \vee z_2))\\
&\hspace{6.4cm}+
\sum \limits_{j= \lfloor n(\theta_1 \wedge \theta_2) \rfloor +1}^{\lfloor n(\theta_1 \vee \theta_2) \rfloor} \mathbb P (\Delta_j^n X \geq z_{I(\theta_1,\theta_2)}) \bigg\} \\
&=
\big\{ (\theta_1 \wedge \theta_2) + O(n^{-1}) \big\} \times \big\{ \nu([z_1 \wedge z_2 , z_1 \vee z_2)) + O(\Delta_n) \big\}  \\
&\hspace{4cm} + \big\{ \left| \theta_1 - \theta_2 \right| + O(n^{-1}) \big\} \times \big\{ \nu([z_{I(\theta_1, \theta_2)}, \infty)) + O(\Delta_n) \big\}
\end{align*}
as $n\to \infty$, where the $O$-terms are uniform in $(\theta_1,z_1),(\theta_2,z_2) \in A_k$ for the same reason as in equation \eqref{Wahrschabschforallieqn}. Thus $\rho_n^2$ converges even uniformly on each $A_k \times A_k$ to $\rho^2$.
Consequently, for  any sequences
$(\theta_1^{\scriptscriptstyle (n)},z_1^{\scriptscriptstyle  (n)})_{n \in \mathbb N},(\theta_2^{\scriptscriptstyle (n)},z_2^{\scriptscriptstyle (n)})_{n \in \mathbb N} \subset A_k$ such that
$\rho(\theta_1^{\scriptscriptstyle (n)}, z_1^{\scriptscriptstyle (n)} ; \theta_2^{\scriptscriptstyle (n)}, z_2^{\scriptscriptstyle (n)}) \rightarrow 0$, it follows
$\rho_n(\theta_1^{\scriptscriptstyle (n)}, z_1^{\scriptscriptstyle (n)} ; \theta_2^{\scriptscriptstyle (n)}, z_2^{\scriptscriptstyle (n)}) \rightarrow 0$.

Finally, $\rho$ is a semimetric: applying first the triangle inequality in $\mathbb R^n$ and then the Minkowski inequality, one sees that each $\rho_n$ satisfies the
triangle inequality. Thus the triangle inequality also holds for $\rho$.
\qed

\subsection{Proof of Corollary \ref{BueVetResVerallg}}
For $k \in \mathbb N$, choose $K=K(k^{-1})$ and $t_0 = t_0(k^{-1})$ as in Lemma \ref{WahrschAbschLem}. Then, for any $(\theta,z) \in A_k$ and for sufficiently large $n$, we have
\begin{align*}
| \Gb_n & (\theta,z) - \tilde \Gb_n(\theta,z) | = \sqrt{k_n} | \mathbb E U_n(\theta,z) - \theta \nu([z, \infty))|  \\
&\leq
\sqrt{k_n} \left | \frac{1}{n} \sum \limits_{j=1}^{\lfloor n\theta \rfloor} \{ \Delta_n^{-1} \Prob(\Delta_j^n X \geq z) - \nu([z, \infty)) \} \right|
+ \sqrt{k_n} \nu([z, \infty)) \left| \frac{\lfloor n\theta \rfloor}{n} - \theta \right|  \\
&{\leq} K \sqrt{k_n} \Delta_n + \nu([k^{-1} , \infty)) \sqrt{\frac{\Delta_n}{n}} \rightarrow 0.
\end{align*}
because of equation \eqref{Wahrschabschforallieqn}. Since the convergence is uniform in $(\theta,z) \in A_k$, we obtain that $d(\Gb_n, \tilde \Gb_n) \to  0$ in probability. Lemma 1.10.2(i) in
\cite{VanWel96} yields the assertion.
\qed

\subsection{Proof of Theorem \ref{TnSchwKonv}}
We are going to use the extended continuous mapping theorem (Theorem 1.11.1 in \citealp{VanWel96}). For $n \in \mathbb N_0$, define
$g_n \colon \mathcal B_{\infty}(\mathbb A) \rightarrow \mathcal B_{\infty}(\mathbb A)$ through
\begin{align*}
g_n(f) &(\theta,z)= f(\theta,z) - \frac{\lfloor n\theta \rfloor}{n} f(1,z) , \quad \text{ for } n \in \mathbb N \\
\intertext{and }
g_0(f) &(\theta,z) = f(\theta,z) - \theta f(1,z) \rbrace.
\end{align*}
Note that $g_n$ is Lipschitz continuous for any $n\in \N_0$.

Obviously, $\Tb_n = g_n(\Gb_n)+ \Exp \Tb_n$ for each $n \in \mathbb N$ and $\Tb=g_0(\Gb)$.
We have
$$\Exp \Tb_n(\theta,z) = \sqrt{k_n} \lambda_n(\theta) \left \{ \frac{n}{\ip{n \theta}} \Exp U_n(\theta,z) - \frac{n}{n- \ip{n \theta}} [\Exp U_n(1,z) - \Exp U_n(\theta,z)]  \right \}$$
and the proof of Corollary \ref{BueVetResVerallg} shows that $\Exp \Tb_n$ converges to $0$ in $\mathcal B_{\infty}(\mathbb A)$.
Thus, by Slutsky's theorem (\citealp{VanWel96}, Example 1.4.7), it suffices to verify $g_n(\Gb_n) \weak g_0(\Gb)$.

Due to Theorem 1.11.1 in \cite{VanWel96} (note that $\Gb$ is separable as it is tight; see Lemma 1.3.2 in
the last-named reference) this weak convergence is valid, if we can show that, for any sequence
$(f_n)_{n \in \mathbb N} \subset \mathcal B_{\infty}(\mathbb A)$ with $f_n \rightarrow f_0$ for some
$f_0 \in \mathcal B_{\infty}(\mathbb A)$,
we have
$$g_n(f_n) \rightarrow g_0(f_0).$$
Let $(f_n)_{n \in \mathbb N}$ be such a sequence with limit point $f_0$. Convergence in
$(\mathcal B_{\infty}(\mathbb A),d)$ is equivalent to uniform convergence on each $A_k$ with $k \in \mathbb N$. The latter is true since\begin{align*}
\| g_n(f_n) - g_0(f_0) \|_{A_k} &= \| f_n(\theta,z) - (\lfloor n\theta \rfloor/n) f_n(1,z) - f_0(\theta,z) + \theta f_0(1,z) \|_{A_k} \\
                                & \leq n^{-1} \| f_0 \|_{A_k} + 2 \| f_n - f_0 \|_{A_k}.
\end{align*}
Obviously, $\Tb$ is a tight, mean-zero Gaussian process. Moreover, from Theorem~\ref{WeakConvThm},
\begin{align*}
\Cov\{ \mathbb T(\theta_1,z_1), \mathbb T(\theta_2,z_2) \} &= H(\theta_1,z_1;\theta_2,z_2) - \theta_1 H(1,z_1;\theta_2,z_2) \\
& \hspace{3cm} - \theta_2 H(\theta_1,z_1;1,z_2) + \theta_1 \theta_2 H(1,z_1;1,z_2) \\															 &= \{ (\theta_1 \wedge \theta_2) - \theta_1 \theta_2 \} \nu([z_1 \vee z_2 , \infty))
\end{align*}
for any $(\theta_1,z_1),(\theta_2,z_2) \in \mathbb A$.
\qed

\subsection{Proof of Proposition \ref{SchwKonvforz0}}
Because of Corollary \ref{BueVetResVerallg} (and the continuous mapping theorem) $U_{1:n}(z_0) = U_n(1,z_0)$ converges to
$\nu([z_0,\infty)) >0$ in probability. Therefore, it follows easily that the random variable $ \{U_n(1,z_0)\}^{-1/2} \mathtt 1_{\lbrace U_n(1,z_0) > 0 \rbrace}$ converges to $\{ \nu([z_0, \infty)) \}^{-1/2}$ in probability.
Hence, by Slutsky's theorem (\citealp{VanWel96}, Example 1.4.7) we obtain
$$V_n^{(z_0)}(\theta) \weak  \frac{1}{\sqrt{\nu([z_0, \infty))}} \mathbb T(\theta,z_0).$$
By Theorem \ref{TnSchwKonv} the process on the right-hand side of this display is a tight mean zero Gaussian with covariance function $k(\theta_1,\theta_2) = \theta_1 \wedge \theta_2 - \theta_1 \theta_2$.
Thus, the law of that process is the law of a standard Brownian bridge on $\ell^{\infty}([0,1])$.
\qed

\subsection{Proof of Theorem \ref{GnVorberKonvThm}}
Due to Lemma \ref{EinschraufTigen} below it suffices to prove conditional weak convergence on $\ell^{\infty}(A_k)$ for any fixed  $k \in \mathbb N$. Recall the triangular array $\lbrace f_{nj}(\omega ;\theta,z) \mid n \geq 1; j=1, \ldots ,n; (\theta,z) \in A_k \rbrace$ consisting of the processes
\[
f_{nj}(\omega ;\theta,z) \defeq k_n^{-1/2} \mathtt 1_{\lbrace j \leq \lfloor n\theta \rfloor \rbrace}
\mathtt 1_{\lbrace \Delta_j^n X \geq z \rbrace}.
\]
Set $\mu_{nj}(\theta,z) \defeq \mathbb E f_{nj}( \cdot; \theta,z)= k_n^{-1/2} \mathtt 1_{\lbrace j \leq \lfloor n\theta \rfloor \rbrace}  \Prob(\Delta_j^n X \ge z)$ and let
$$
\hat \mu_{nj}(\theta,z)
\defeq
\hat \mu_{nj}(\omega; \theta,z)
\defeq
k_n^{-1/2} \mathtt 1_{\lbrace j \leq \lfloor n\theta \rfloor \rbrace}  \eta_n(z)
$$
be an estimator for $\mu_{nj}(\theta,z)$.
Then,  $\hat \Gb_n$ can be written as
\[
\hat \Gb_n(\theta,z) = \hat \Gb_n(\omega; \theta,z) = \sum \limits_{j=1}^n \xi_j  \{ f_{nj}(\omega; \theta,z) - \hat \mu_{nj}(\omega; \theta,z) \}.
\]
Due to Theorem 3 in \cite{Kos03} the proof is complete, if we show the following properties for the triangular array
$\lbrace \hat \mu_{nj}(\omega ;\theta,z) \mid n \geq 1; j=1, \ldots ,n; (\theta,z) \in A_k \rbrace$:
\medskip
\begin{compactenum}[(i)]
\item $\lbrace \hat \mu_{nj} \rbrace$ is almost measurable Suslin. \label{Liste1}
\item $\sup \limits_{(\theta,z) \in A_k} \sum \limits_{j=1}^n \{ \hat \mu_{nj}(\omega; \theta,z) - \mu_{nj}(\theta,z) \}^2
      \pto  0.$ \label{Liste2}
\item The triangular array $\lbrace \hat \mu_{nj} \rbrace$ is manageable with envelopes $\lbrace \hat F_{nj} \rbrace$ given
      through $ \hat F_{nj} (\omega) \defeq k_n^{-1/2} n^{-1} \sum_{i=1}^n
			\mathtt 1_{\lbrace \Delta_i^n X \geq k^{-1} \rbrace} $. \label{Liste3}
\item There exists a constant $M < \infty$ such that $M \vee \sum \limits_{j=1}^n \hat F_{nj}^2
      \pto  M$. \label{Liste4}
\end{compactenum}

\smallskip
\noindent
\textit{Proof of  \eqref{Liste1}.}
As in the proof of (1) in Theorem \ref{WeakConvThm}, it suffices to verify that the triangular array
$\lbrace \hat \mu_{nj} \rbrace$ is separable. This can be seen by taking $S_n \defeq A_k \cap \mathbb Q^2$ again.

\smallskip
\noindent
\textit{Proof of  \eqref{Liste2}.}
We have
\begin{align*}
\sup \limits_{(\theta,z) \in A_k} \sum_{j=1}^n \{ \hat \mu&_{nj}(\omega; \theta,z) - \mu_{nj}(\theta,z) \} ^2 \\
&= \sup \limits_{z \ge 1/k} n^{-3} \Delta_n^{-1} \sum \limits_{j = 1}^n \bigg [ \sum \limits_{i=1}^n \{ \mathtt 1_{\lbrace \Delta_i^n X \ge z \rbrace} - \Prob (\Delta_j^n X \ge z) \} \bigg ]^2 \\
&= n^{-1} \sup \limits_{z \ge 1/k} \{ \Gb_n(1,z) \}^2 + O_{\Prob}(\Delta_n^{2}),
\end{align*}
where the final approximation error is a consequence of equation \eqref{Wahrschabschforallieqn} in the proof of Theo\-rem~\ref{WeakConvThm}. The last quantity in the above display converges to $0$ in probability by Theo\-rem~\ref{WeakConvThm}.

\smallskip
\noindent
\textit{Proof of  \eqref{Liste3}.}
In the proof of Theorem \ref{WeakConvThm} we have already shown that the triangular array
$$
\lbrace \tilde g_{nj} (\theta) \defeq \tilde g_{nj} (\omega;\theta) \defeq \mathtt 1_{\lbrace j \leq \lfloor n\theta \rfloor \rbrace} \mid n \in \mathbb N; j=1, \ldots, n; \theta \in [0,1]\rbrace
$$
is manageable with envelopes $\lbrace \tilde G_{nj} (\omega) \stackrel{\mathrm{def}}\equiv 1 \mid n \in \mathbb N; j=1, \ldots,n \rbrace$.
Therefore, due to Theorem 11.17 in \cite{Kos08}, it suffices to prove that the triangular array
$$
\left\{ \tilde h_{nj} (\omega;z)
\defeq
\frac{1}{n \sqrt{k_n}} \sum \limits_{i=1}^n \mathtt 1_{\lbrace \Delta_i^n X
\geq z \rbrace} \mid n \in \mathbb N; j=1, \ldots, n; z \in [k^{-1}, \infty) \right\}
$$
is manageable with envelopes $\lbrace \hat F_{nj} (\omega) \mid n \in \mathbb N; j=1, \ldots,n \rbrace$. But $\tilde h_{nj} (\omega;z)$ does not depend on $j$ at all, such that every projection of
$\mathcal H_{n \omega} \defeq \lbrace (\tilde h_{n1}(\omega;z), \ldots, \tilde h_{nn}(\omega; z)) \mid z \geq k^{-1} \rbrace$ onto two coordinates lies in the straight line $\lbrace (x,y) \in \mathbb R^2 \mid x=y \rbrace$. Consequently, the set
$\mathcal H_{n \omega}$ has a pseudodimension of at most $1$ (Definition 4.3 in \citealp{Pol90}) and is bounded. Hence, the same arguments as in the proof of Theorem \ref{WeakConvThm} show the desired manageability.

\smallskip
\noindent
\textit{Proof of  \eqref{Liste4}.}
A straight forward calculation yields
$$
\mathbb E \left\{ \sum \limits_{j=1}^n \hat F_{nj}^2 \right\}
 = n^{-2} \Delta_n^{-1} \sum \limits_{i_1 =1}^n \sum \limits_{i_2 =1}^n \mathbb E \left\{ \mathtt 1_{\lbrace \Delta_{i_1}^n X \ge k^{-1} \rbrace}
\mathtt 1_{\lbrace \Delta_{i_2}^n X \ge k^{-1} \rbrace} \right\} = O(\Delta_n).
$$
Here we used equation \eqref{Wahrschabschforallieqn} again and the fact, that the increments of $X$ are independent since we assume deterministic characteristics. Thus $\sum \limits_{j=1}^n \hat F_{nj}^2$ is $o_\Prob(1)$.
\qed

\subsection{Proof of Theorem \ref{BootstrTeststThm}}
Again by Lemma \ref{EinschraufTigen} it suffices to prove the convergence in the spaces $\ell^{\infty}(A_k)$. Let therefore $k \in \mathbb N$ be fixed for the rest of the proof.

By definition, $\hat \Tb_n = g_n(\hat \Gb_n)$ in $\ell^\infty(A_k)$, with $g_n$ defined in the proof of Theorem \ref{TnSchwKonv}. Now,
$(\ell^{\infty}(A_k), \| \cdot \|_{A_k})$ is a Banach space and the mapping $g_0 \colon \ell^{\infty}(A_k) \rightarrow \ell^{\infty}(A_k)$ defined in the proof of Theorem \ref{TnSchwKonv} is Lipschitz continuous. Hence, Proposition~10.7(i) in \cite{Kos08} yields the convergence
$$
g_0( \hat \Gb_n) \weakP g_0( \mathbb G) = \mathbb T
$$
in $\ell^{\infty}(A_k)$.
Furthermore, due to the definition of the mappings $g_n, g_0$ and the definition of the process $\hat \Gb_n$ we obtain that
$$
\| g_n(\hat \Gb_n) - g_0(\hat \Gb_n) \|_{A_k}
\leq
\frac{1}{n} \sup \limits_{z \geq \frac{1}{k}} | \hat \Gb_n(1,z) |
$$
By Theorem~\ref{GnVorberKonvThm}, the right-hand side converges to $0$ in probability.
Another application of Lemma~\ref{NaheaequivKonvLemma} shows that
$\Tb_n = g_n(\hat \Gb_n)  \weakP g_0 (\mathbb G) = \mathbb T$
as asserted.
\qed

\subsection{Proof of Proposition~\ref{CorrKSCP}}

The assertion that $\lim_{n \rightarrow \infty} \mathbb P( V_n^{(z_0)} \geq q^K_{1- \alpha}) = \alpha$ under $\textbf H_0$ is a simple consequence of Proposition \ref{SchwKonvforz0} and the fact that the KS-distribution has a continuous cumulative distribution function.

With respect to the assertion regarding $W_n^{\scriptscriptstyle (z_0)}$ note that, under $\textbf H_0$, Proposition~\ref{JointConvProp} and the continuous mapping theorem imply that, for any fixed $B \in \mathbb N$,
\[
(W_n^{(z_0)}, \hat W_{n, \xi^{(1)}}^{(z_0)}, \ldots, \hat W_{n, \xi^{(B)}}^{(z_0)}) \weak  (W^{(z_0)}, W^{(z_0),(1)}, \ldots, W^{(z_0),(B)})
\]
in $\mathbb R^{B+1}$, where $W^{(z_0)} \defeq \sup_{\theta \in [0,1]} |\mathbb T(\theta,z_0) |$ with the limit process $\mathbb T$ of Theorem~\ref{TnSchwKonv} and where  $W^{(z_0),(1)}, \ldots,W^{(z_0),(B)}$ are independent copies of $W^{(z_0)}$. According to the corollary to Proposition 3 in \cite{Lif84}, $W^{(z_0)}$ has a continuous c.d.f.\ under $\textbf H_0$. Thus, Proposition F.1 in the supplement to \cite{BueKoj14} implies that
\[
\lim \limits_{B \rightarrow \infty} \lim \limits_{n \rightarrow \infty} \mathbb P\{ W_n^{(z_0)} \geq \hat q_{1 - \alpha}^{(B)}(W_n^{(z_0)})\} = \alpha
\]
for all $\alpha \in (0,1)$, as asserted. Observing that, under $\textbf H_0$ and for $\eps > 0$ with $\nu([\eps, \infty)) > 0$, the distribution of $T^{(\eps)}$ has a continuous c.d.f., essentially the same reasoning also implies that
\[
\lim_{B \rightarrow \infty} \lim_{n \rightarrow \infty} \mathbb P\{ T_n^{(\eps)} \geq \hat q_{1- \alpha}^{(B)}(T_n^{(\eps)}) \} = \alpha
\]
for all $\alpha \in (0,1)$. \qed

\subsection{Proof of Proposition~\ref{CorConsi}}
In order to prove consistency of the CP-Test, choose $\eps  >0$ as in Proposition~\ref{prop:h1unbounded} such that $\lim_{n \rightarrow \infty} \mathbb P( T_n^{\scriptscriptstyle (\eps)} \geq K) =1$ for any $K>0$.
By Proposition \ref{prop:bootbound}, for given $\delta >0$ and fixed $B \in \mathbb N$, we may  choose $K_0>0$ such that
$$\sup \limits_{n \in \mathbb N} \mathbb P\Big(\max \limits_{b = 1, \ldots, B} \hat T_{n, \xi^{(b)}}^{(\eps)} > K_0\Big) \leq \frac{\delta}{2}.$$
For this $K_0$, we can now take $N \in \mathbb N$ such that
$$\mathbb P(T_n^{(\eps)} \geq K_0) \geq 1 - \frac{\delta}{2}$$
holds for all $n \geq N$. Then, for any $n \geq N$,
\begin{align*}
1 - \delta
&\leq
\mathbb P(T_n^{(\eps)} \geq K_0) - \mathbb P\Big(\max \limits_{b = 1, \ldots, B} \hat T_{n, \xi^{(b)}}^{(\eps)} > K_0\Big) \\
&\leq
\mathbb P\Big(T_n^{(\eps)} \geq K_0 , \max \limits_{b = 1, \ldots, B} \hat T_{n, \xi^{(b)}}^{(\eps)} \leq K_0\Big) \\
&\leq
\mathbb P\big\{ T_n^{(\eps)} \geq \hat q_{1 - \alpha}^{(B)}(T_n^{(\eps)}) \big\}.
\end{align*}
This proves the assertion for the CP-Test, and the claim for KSCP-Test2 follows along the same lines.
The assertion for KSCP-Test1 is a direct consequence of Proposition \ref{prop:h1unbounded}.
\qed

\subsection{Proof of Proposition \ref{prop:tnh1}}
Let $X^{\scriptscriptstyle (1)}(n)$ and $X^{\scriptscriptstyle (2)}(n)$ denote two independent It\=o semimartingales with characteristics $(b^{\scriptscriptstyle(n)}_t, \sigma^{\scriptscriptstyle(n)}_t, \nu_1)$ and $(b^{\scriptscriptstyle(n)}_t, \sigma^{\scriptscriptstyle(n)}_t, \nu_2)$, respectively.
For $n \in \mathbb N$ and $j=0, \dots, n$, set $Y_j(n) = X_{j\Delta_n}^{\scriptscriptstyle (1)}(n)$ and $Z_j(n)=X_{j\Delta_n}^{\scriptscriptstyle (2)}(n)$.
Let $U_n^{\scriptscriptstyle(1)}$ and $U_n^{\scriptscriptstyle(2)}$ denote the quantity defined in \eqref{UnVerallgSchaetz}, based on the observations $Y_j(n)$ and $Z_j(n)$, respectively, instead on $X_{j\Delta_n}$. 
Moreover, define a random element $S_n$ with values in $\mathcal B_{\infty}(\mathbb A)$ through
\begin{multline*}
S_n(\theta,z) 
:= 
\frac{n - \lfloor n\theta \rfloor}{n} U_n^{(1)}(\theta,z) - \frac{\lfloor n\theta \rfloor}{n} \{ U_n^{(1)}(\theta_0,z) - U_n^{(1)}(\theta,z) \}   \\
- \frac{\lfloor n\theta \rfloor}{n}  \{ U_n^{(2)}(1,z) - U_n^{(2)}(\theta_0,z) \},
\end{multline*}
for $(\theta,z) \in \mathbb A$ with $\theta \le \theta_0$, whereas for $(\theta,z) \in \mathbb A$ with $\theta \ge \theta_0$,
\begin{multline*}
S_n(\theta,z) := \frac{n - \lfloor n\theta \rfloor}{n} U_n^{(1)}(\theta_0,z) + \frac{n- \lfloor n\theta \rfloor}{n} \{ U_n^{(2)}(\theta,z) - U_n^{(2)}(\theta_0,z) \} \\
-  \frac{\lfloor n\theta \rfloor}{n} \{ U_n^{(2)}(1,z) - U_n^{(2)}(\theta,z) \} .
\end{multline*}
According to Theorem II.4.15 in \cite{JacShi02}, we have the distributional equality
\begin{multline*}
\big(\Delta_1^n X(n) , \ldots, \Delta_{\ip{n \theta_0}}^n X(n) , \Delta_{\ip{n \theta_0}+1}^n X(n) , \ldots, \Delta_n^n X(n) \big)  \\ 
\stackrel{\mathcal D}= 
\big(\Delta_1^n X^{(1)}(n), \ldots, \Delta_{\ip{n \theta_0}}^n X^{(1)}(n), \Delta_{\ip{n \theta_0}+1}^n X^{(2)}(n), \ldots, \Delta_n^n X^{(2)}(n) \big).
\end{multline*}
Hence, for any $(\theta_1,z_1), \ldots, (\theta_p,z_p) \in \mathbb A$ and $p \in \mathbb N$, we also have that
\[
\big (k_n^{-1/2} \Tb_n(\theta_1,z_1), \ldots, k_n^{-1/2} \Tb_n(\theta_g,z_g) \big ) 
\stackrel{\mathcal D}=  \big(S_n(\theta_1,z_1), \ldots, S_n(\theta_g,z_g) \big).
\]

By Theorem 1.6.1 in \cite{VanWel96}, we have to show uniform convergence of $k_n^{\scriptscriptstyle-1/2}\Tb_n$ to $T$ on any $A_k$ with $k\in \N$, in probability.  Now, from the previous display, and from the fact that the function $T$ is continuous in $(\theta,z)$ and that the functions $\Tb_n(\theta,z)$ depend only through $\ip{n \theta}$ on $\theta$ and are left-continuous in $z$, we immediately get that
\begin{align*}
\sup \limits_{(\theta, z)\in A_k} \left| k_n^{-1/2} \Tb_n(\theta,z) - T(\theta,z) \right|
&=
\sup \limits_{(\theta, z)\in A_k \cap \mathbb Q^2} \left| k_n^{-1/2} \Tb_n(\theta,z) - T(\theta,z) \right| \\
&\stackrel{\mathcal D}= \sup \limits_{(\theta, z)\in A_k \cap \mathbb Q^2} \left| S_n(\theta,z) - T(\theta,z) \right|
\end{align*}
This expression is in fact $o_\Prob(1)$ as a consequence of Corollary \ref{BueVetResVerallg} and the continuous mapping theorem. 
Note that the proofs of Lemma \ref{WahrschAbschLem}, Theorem \ref{WeakConvThm} and Corollary~\ref{BueVetResVerallg} show that Corollary \ref{BueVetResVerallg} is in fact applicable in this setup, because the characteristics $b^{\scriptscriptstyle (n)}_t$ and $\sigma^{\scriptscriptstyle (n)}_t$ have a uniform bound in $n \in \mathbb N$ and the resulting constants of Lemma \ref{WahrschAbschLem} depend only on the bound of the characteristics and on~$\delta$.
\qed

\subsection{Proof of Proposition \ref{CorConsiLoc}}
Under $\mathbf H_1$, choose $\eps >0$ such that there exists a $z_0 \geq \eps$ with $\nu_1(z_0) \neq \nu_2(z_0)$. Then, according to Proposition~\ref{prop:tnh1} and the continuous mapping theorem, the random functions
$\theta \mapsto \sup_{z \geq \eps} |k_n^{\scriptscriptstyle -1/2} \Tb_n(\theta,z)|$ converge weakly in $\ell^{\infty}([0,1])$ to the continuous function $\theta \mapsto \sup_{z \geq \eps} |T(\theta,z)|$, which has a unique maximum at $\theta_0$.

Similarly, under $\textbf H_1^{\scriptscriptstyle (z_0)}$, the random functions $\theta \mapsto |k_n^{\scriptscriptstyle -1/2} \Tb_n(\theta,z_0)|$ converge weakly in $\ell^{\infty}([0,1])$ to the continuous function $\theta \mapsto |T(\theta,z_0)|$, which also has a unique maximum in $\theta_0$.

Thus, the asserted convergences follow from the argmax-continuous mapping theorem (Theorem 2.7 in \citealp{KimPol90}).
\qed

\subsection{Additional auxiliary results}

The following two auxiliary results are needed for validating the bootstrap procedures defined in Section~\ref{sec:boot}. The first lemma is proved in \cite{Buc11}, Lemma~A.1.

\begin{lemma}
\label{NaheaequivKonvLemma}
Consider two bootstrapped statistics $\hat G_n = \hat G_n(X_1, \ldots, X_n, \xi_1, \ldots, \xi_n)$ and $\hat H_n = \hat H_n(X_1, \ldots, X_n, \xi_1, \ldots, \xi_n)$ in a metric space $(\mathbb D,d)$ with
$d(\hat G_n, \hat H_n) \pto 0$. Then, for a tight Borel measurable process $G$ in $\mathbb D$, we have
$\hat G_n \weakP G$ if and only if
$\hat H_n \weakP G$.
\end{lemma}

For the second auxiliary lemma, let $T_1 \subset T_2 \subset \ldots$ be arbitrary sets and set $T \defeq \bigcup_{k=1}^{\infty} T_k$. Let
$(\mathcal B_{\infty}(T), d)$ be defined as the complete metric space of all real-valued functions on $T$ that are bounded on each
$T_k$, equipped with the metric
$$d(f_1,f_2) = \sum \limits_{k=1}^{\infty} 2^{-k} (\| f_1 - f_2 \|_{T_k} \wedge 1),$$
where $\| \cdot \|_{T_k}$ denotes  the sup-norm on $T_k$ \citep[][Chapter 1.6]{VanWel96}. Bootstrap variables on such spaces  converge weakly conditionally in probability if and only if the same holds true in $(\ell^\infty(T_k), \| \cdot \|_{T_k})$ for all
$k \in \mathbb N$.

\begin{lemma}
\label{EinschraufTigen}
Let $\hat G_n = \hat G_n(X_1, \ldots, X_n, \xi_1, \ldots, \xi_n)$ be a bootstrapped statistic with values in
$\mathcal B_{\infty}(T)$ and let $G$ be a tight Borel measurable process taking values in $\mathcal B_{\infty}(T)$.
Then, $\hat G_n \weakP G$ in $(\mathcal B_{\infty}(T),d)$ if and only if
$\hat G_n \weakP G$ in $(\ell^{\infty}(T_k), \|\cdot\|_{T_k})$ for all $k\in\N$.
\end{lemma}

The proof of this lemma can be found in \cite{Buc11}, Lemma A.5, for a special choice of the $T_k$. The proof, however, is independent of this choice.

The proof of Proposition~\ref{CorrKSCP} is based on the following auxiliary result, establishing unconditional weak convergence of the vector of processes
$(\Tb_n, \hat \Tb_{n, \xi^{(1)}}, \ldots, \hat \Tb_{n, \xi^{(B)}})$.

\begin{proposition}
\label{JointConvProp}
Suppose the conditions from Theorem~\ref{GnVorberKonvThm} are met. Then, under $\mathbf H_0$, for all $B \in \mathbb N$, we have
\begin{align*}
(\Tb_n, \hat \Tb_{n, \xi^{(1)}}, \ldots, \hat \Tb_{n, \xi^{(B)}}) \weak  (\mathbb T ,  \mathbb T^{(1)}, \ldots, \mathbb T^{(B)})
\end{align*}
in $(\mathcal B_{\infty}(\mathbb A),d)^{B+1}$, where $\weak $ denotes (unconditional) weak convergence (with respect to the probability measure $\mathbb P$), and where $\mathbb T^{(1)}, \ldots, \mathbb T^{(B)}$ are independent copies of $\mathbb T$.
\end{proposition}

\noindent
\textit{Proof.}
We are going to apply Corollary 1.4.5 in \cite{VanWel96}. Therefore, let $f^{(0)},f^{(1)}, \ldots, f^{(B)} \in \text{BL}_1(\mathcal B_{\infty}(\mathbb A))$. Since
$\Tb_n, \hat \Tb_{n, \xi^{(1)}}, \ldots, \hat \Tb_{n, \xi^{(B)}}$ are independent conditional on the data, we have
\begin{multline*}
\mathbb E_{\xi}  \big \{  f^{(0)}(T_n) \cdot f^{(1)}(\hat T_{n, \xi^{(1)}}) \cdot \ldots \cdot f^{(B)}(\hat T_{n, \xi^{(B)}}) \big \} \\
=
 f^{(0)}(T_n) \cdot \mathbb E_{\xi} f^{(1)}(\hat T_{n, \xi^{(1)}}) \cdot \ldots \cdot  \mathbb E_{\xi}f^{(B)}(\hat T_{n, \xi^{(B)}}) =: S_n.
\end{multline*}
By Definition \ref{ConvcondDataDef} and Theorem \ref{BootstrTeststThm},  $\mathbb E_{\xi} f^{(b)}(\hat T_{n, \xi^{(b)}})$ converges in outer probability to $\mathbb E(f^{(b)}(\mathbb T^{(b)})) =: c_b$ for each $b \in \lbrace 1, \ldots, B \rbrace$. Therefore,
$$S_n \weak c_1 \cdot \ldots \cdot c_B \cdot f^{(0)}(\mathbb T) =: \mathcal S$$
by using  the continuous mapping theorem, Slutsky's Lemma and Lemma 1.10.2 in \cite{VanWel96} several times.

Choose an $M>0$ with $|S_n| \vee | \mathcal S | \leq M$ for all $\omega \in \Omega$, $n \in \mathbb N$ and let $g \colon \mathbb R \longrightarrow \mathbb R$ be a bounded and continuous function with $g(x) = x$ on $[-M,M]$. Then
\begin{align}
\label{SchwKonvBootGl1}
& \hspace{-1.5cm} \mathbb E_X^{\ast} \Big[ \mathbb E_{\xi}^{\ast}  \big\{ f^{(0)}(\Tb_n) \cdot f^{(1)}(\hat \Tb_{n, \xi^{(1)}}) \cdot \ldots \cdot f^{(B)}(\hat \Tb_{n, \xi^{(B)}}) \big\} \Big] \nonumber \\
&=
\mathbb E_X^{\ast} S_n
=
\mathbb E_X^{\ast} g(S_n)  \nonumber
\, \stackrel{(1)}{\longrightarrow}  \,
\mathbb E(g(\mathcal S)) = \mathbb E \mathcal S \nonumber \\
&\hspace{1.5cm} \stackrel{(2)}=
\mathbb E \big\{ f^{(0)}(\mathbb T) \cdot f^{(1)}(\mathbb T^{(1)}) \cdot \ldots \cdot f^{(B)}(\mathbb T^{(B)}) \big\}.
\end{align}
Note that $(1)$ uses the fact that a coordinate projection on a product probability space is perfect (Lemma 1.2.5 in \citealp{VanWel96}). Moreover, $(2)$ holds because the limit processes are independent.

By Theorem \ref{TnSchwKonv}, Remark \ref{rem:condweak}(ii), Theorem \ref{BootstrTeststThm} and Lemma 1.3.8 and Lemma 1.4.4 in \cite{VanWel96} the vector of processes $(\Tb_n, \hat \Tb_{n, \xi^{(1)}}, \ldots, \hat \Tb_{n, \xi^{(B)}})$ is (jointly) asymptotically measurable.
Consequently, Equation \eqref{SchwKonvBootGl1}, Fubini's theorem (Lemma 1.2.6 in \citealp{VanWel96}) and Corollary 1.4.5 in \cite{VanWel96} yield the desired weak convergence. Note that the limit process $(\Tb, \Tb^{(1)}, \ldots, \Tb^{(B)})$ is separable because it is tight (Lemma 1.3.2 in the previously mentioned reference).
\qed

\begin{proposition}
\label{prop:h1unbounded}
Suppose the sampling scheme meets the conditions from Corollary~\ref{BueVetResVerallg}. Then,
under $\mathbf H_1$, there exists an $\eps > 0$ such that, for all $K>0$,
$$\lim \limits_{n \rightarrow \infty} \mathbb P( T_n^{(\eps)} \geq K) =1.$$
If $\mathbf H_1^{(z_0)}$ is true, the same assertion holds for $V_n^{(z_0)}$ and $W_n^{(z_0)}$.
\end{proposition}

\noindent
\textit{Proof.} Choose $\eps >0$ such that there exists a $\hat z \ge \eps$ with $\nu_1(\hat z) \neq \nu_2(\hat z)$.
Then $c \defeq \sup_{\theta \in [0,1]} \sup_{z \ge \eps} \left| T(\theta, z) \right| \in (0,\infty)$, with the function $T$ defined in Proposition~\ref{prop:tnh1}. But Proposition \ref{prop:tnh1} and the continuous mapping theorem show that $k_n^{\scriptscriptstyle -1/2} T_n^{\scriptscriptstyle (\eps)} = c + o_{\Prob}(1)$ and this yields the assertion for $T_n^{\scriptscriptstyle (\eps)}$.

The same argument implies the claim for $W_n^{\scriptscriptstyle (z_0)}$, using the fact that $\nu_1(z_0) \neq \nu_2(z_0)$ and consequently $\sup_{ \scriptscriptstyle \theta \in [0,1]} \left| T(\theta, z_0) \right| >0$ under $\mathbf H_1^{\scriptscriptstyle (z_0)}$.

Finally, let us prove the claim for $V_n^{\scriptscriptstyle (z_0)}$. As in the proof of Proposition~\ref{prop:tnh1}, let $X^{\scriptscriptstyle (1)}(n)$ and $X^{\scriptscriptstyle (2)}(n)$ be independent It\=o semimartingales with characteristics $(b^{\scriptscriptstyle (n)}_t, \sigma^{\scriptscriptstyle (n)}_t, \nu_1)$ and $(b^{\scriptscriptstyle (n)}_t, \sigma^{\scriptscriptstyle (n)}_t, \nu_2)$, respectively.  For $n \in \mathbb N$ and $j=0, \dots, n$, set $Y_j(n) = X_{j\Delta_n}^{\scriptscriptstyle (1)}(n)$ and $Z_j(n)=X_{j\Delta_n}^{\scriptscriptstyle (2)}(n)$.
Let $U_n^{\scriptscriptstyle(1)}$ and $U_n^{\scriptscriptstyle(2)}$ denote the quantity defined in \eqref{UnVerallgSchaetz}, based on the observations $Y_j(n)$ and $Z_j(n)$, respectively, instead on $X_{j\Delta_n}$.

Then the quantities $V_n^{\scriptscriptstyle(z_0)}$ and $W_n^{\scriptscriptstyle(z_0)}$ differ only by a factor $A_n^{\scriptscriptstyle-1/2} \mathtt 1_{\lbrace A_n > 0 \rbrace}$, with $A_n$ being equal in distribution to (Theorem II.4.15 in \citealp{JacShi02})
$$U^{(1)}_n(\theta_0,z_0) + U_n^{(2)}(1,z_0) - U_n^{(2)}(\theta_0, z_0).$$
This expression converges to $\theta_0 \nu_1(z_0) + (1- \theta_0) \nu_2(z_0) > 0$, in probability, which in turn implies the assertion regarding $V_n^{(z_0)}$.
\qed


\begin{proposition}
\label{prop:bootbound}
Suppose the sampling scheme meets the conditions from Corollary~\ref{BueVetResVerallg}. Then,
under $\mathbf H_1$, for all $\eps > 0$ and all $b \in \lbrace 1, \ldots , B \rbrace$,
$$\hat T_{n, \xi^{(b)}}^{(\eps)} = \text{ O}_{\mathbb P} (1), ~~\text{ that is } \quad \lim \limits_{K \rightarrow \infty} \limsup \limits_{n \rightarrow \infty} \mathbb P (\hat T_{n, \xi^{(b)}}^{(\eps)} > K) = 0.$$
Moreover, under $\mathbf H_1^{(z_0)}$, for all $b \in \lbrace 1, \ldots , B \rbrace$,
$$\hat{W}_{n, \xi^{(b)}}^{(z_0)} = \text{ O}_{\mathbb P} (1), ~~\text{ that is } \quad \lim \limits_{K \rightarrow \infty} \limsup \limits_{n \rightarrow \infty} \mathbb P (\hat{W}_{n, \xi^{(b)}}^{(z_0)} > K) = 0.$$
\end{proposition}

\noindent
\textit{Proof.}
Since the results are independent of $b$, we omit this index throughout the proof. Also note that, for both assertions, it suffices to show that, for any $k\in\N$, $\sup_{\theta \in [0,1]} \sup_{z \ge 1/k} | \hat \Gb_n(\theta,z) |=O_\Prob(1)$ under $\mathbf H_1$.

For $n \in \mathbb N$ and $j=0, \dots, n$, let $Y_j(n) = X_{j\Delta_n}^{\scriptscriptstyle (1)}(n)$ and $Z_j(n)=X_{j\Delta_n}^{\scriptscriptstyle (2)}(n)$ be defined as in the proof of Proposition \ref{prop:tnh1}.
Let $U_n^{\scriptscriptstyle(1)}$, $\eta_n^{\scriptscriptstyle (1)}$ and $U_n^{\scriptscriptstyle(2)}$, $\eta_n^{\scriptscriptstyle (2)}$ denote the corresponding quantities, based on the observations $Y_j(n)$ and $Z_j(n)$, respectively, instead on $X_{j\Delta_n}$.

Then, for $\theta \le \theta_0$, we can write $\hat \Gb_n(\theta,z)$ as
\[
\frac{1}{\sqrt{k_n}} \sum_{j=1}^{\ip{n\theta} } \xi_j  \{  \mathtt 1_{\lbrace \Delta_j^n Y \geq z \rbrace} - \eta_n^{(1)}(z) \} \\
+
\left\{ \frac{1}{\sqrt n}  \sum_{j=1}^{\ip{n\theta} } \xi_j  \right\} \times \left \{ \Delta_n^{-1/2} \big( \eta_n^{(1)} - \eta_n \big) \right\} .
\]
The first term of this display is $O_\Prob(1)$, uniformly in $\theta \le \theta_0$ and $z \ge 1/k$, by Theorem~\ref{GnVorberKonvThm} and Remark \ref{rem:condweak} (ii). By the classical Donsker theorem, the first term in curly brackets on the right-hand side is also $O_\Prob(1)$ uniformly in $\theta \le \theta_0$. 
The quantity $\Delta_n^{\scriptscriptstyle -1/2} \eta_n^{\scriptscriptstyle (1)}(z) = \Delta_n^{\scriptscriptstyle 1/2} U^{\scriptscriptstyle (1)}_n(1,z)$ is $o_\Prob(1)$ by Corollary~\ref{BueVetResVerallg}. Finally, the same argument as in the proof of Proposition \ref{prop:tnh1} yields
\[
\Delta_n^{-1/2} \sup \limits_{z \ge 1/k} \left| \eta_n(z) \right| = \sqrt {\Delta_n} \sup \limits_{z \ge 1/k} \left| U_n^{(1)} (\theta_0,z)  + U_n^{(2)} (1,z) -U_n^{(2)}(\theta_0,z) \right| = o_{\Prob}(1).
\]
To conclude,
\[
\sup_{\theta \le \theta_0} \sup_{z \ge 1/k} | \hat \Gb(\theta,z) | =O_\Prob(1).
\]
The supremum over $\theta > \theta_0$ and  $z \ge 1/k$ can be treated similarly.
\qed

\bigskip
\noindent
\textbf{Acknowledgements.}
This work has been supported by the Collaborative Research Center ``Statistical modeling of nonlinear dynamic processes'' (SFB 823, Teilprojekt A1, A7, C1) 
of the German Research Foundation (DFG) which is gratefully acknowledged.

\bibliographystyle{chicago}
\bibliography{biblio}
\end{document}